\documentclass{aa}
\usepackage{graphicx}
\usepackage[varg]{txfonts}
\usepackage{longtable}
\usepackage{lscape}

\begin{document}

\title{Activity time series of old stars from late F to early K IV. Diagnosis from photometry}
%   \subtitle{}

\titlerunning{}

\author{N. Meunier \inst{1}, A.-M. Lagrange \inst{1} 
  }
\authorrunning{Meunier et al.}

\institute{
Univ. Grenoble Alpes, CNRS, IPAG, F-38000 Grenoble, France\\
\email{nadege.meunier@univ-grenoble-alpes.fr}
     }

\offprints{N. Meunier}

\date{Received ; Accepted}

\abstract{A number of high-precision time series have recently become available for many stars as a result of data from CoRoT, Kepler, and TESS. These data have been widely used to study stellar activity. Photometry provides information that is integrated over the stellar disk. Therefore, there are many degeneracies  between spots and plages or sizes and contrasts. In addition, it is important to relate activity indicators, derived from photometric light curves, to other indicators (LogR'$_{\rm{HK}}$ and radial velocities).}
{Our aim is to understand how to relate photometric variability to physical parameters in order to help the interpretation of these observations. }
{We used a large number of synthetic time series of brightness variations for old main sequence stars within the F6-K4 range.\ Simultaneously, we computed using consistent modeling for radial velocity, astrometry, and chromospheric emission. We analyzed these time series to study the effect of the star  spectral type on brightness variability,  the relationship between brightness variability and chromospheric emission, and the interpretation of brightness variability as a function of spot and plage properties. We then studied spot-dominated or plage-dominated regimes. }
{We find that within our range of activity levels, the brightness variability increases toward low-mass stars, as suggested by Kepler results. However, many elements can create an interpretation bias. Brightness variability roughly correlates to LogR'$_{\rm{HK}}$  level. There is, however, a large dispersion in this relationship, mostly caused by spot contrast and inclination. It is also directly related to the number of structures, and we show that it can not be interpreted solely in terms of spot sizes. Finally, a detailed analysis of its relation with LogR'$_{\rm{HK}}$ shows that in the activity range of old main-sequence stars, we can obtain both spot or plage dominated regimes, as was shown by observations in previous works. The same star can also be observed in both regimes depending on inclination. Furthermore, only strong correlations between chromospheric emission and brightness variability are significant.}
{Our realistic time series proves to be extremely useful when interpreting observations and understanding their limitations, most notably in terms of activity interpretation. Inclination is crucial and affects many properties, such as  amplitudes and the respective  role of spots and plages. }


\maketitle

%%%%%%%%%%%%%%%%%%%%%%%%%%%%%%%%%%%%%%%%%%%%%%%%%%%%%
%%%%%%%%%%%%%%%%%%%%%%%%%%%%%%%%%%%%%%%%%%%%%%%%%%%%%
%-----------------------------------
%-----------------------------------
\section{Introduction}
%-----------------------------------
%-----------------------------------

Although the Sun's activity is very well characterized, this is not so for the activity of other stars. Stellar activity is however ubiquitous and has been observed using many complementary techniques (e.g., Zeeman-Doppler imaging, brightness variability in various wavelength ranges, X-ray emission, and chromospheric emission). For instance, most measurements are based on the interpretation of disk-integrated observables for relatively old solar-type stars with low rotation rates. Interpreting the photometric variability in terms of the properties of spots and plages has been done in the past \cite[e.g.,][]{lanza09,mosser09,kipping12,juvan18}. However, it is difficult in the case of solar type stars with complex activity patterns \cite[][]{lanza07,bonomo08}.\ This occurs because of strong degeneracies in the structure properties between size and contrast, for example. 
Simulating realistic complex activity patterns in stars, and the resulting time series of observables, is therefore crucial when interpreting these observations and determining the limits of these interpretations. It is also useful to test analysis methods on time series in which the parameters are controlled, such as when testing the rotation rate measurement \cite[e.g., as done by][]{reinhold13,mcquillan13b,arkhypov15,reinhold17}. This is particularly important since a wealth of high-precision photometry time series have recently become  available. This was first the case with CoRoT and Kepler, and more recently with TESS for a very large number of stars. These data have been widely used to study stellar activity \cite[e.g.,][]{basri10,basri11,reinhold13,reinhold13b,nielsen13,lanza14,mcquillan14,arkhypov15,ferreiralopes15,he15,mehrabi17,reinhold17,reinhold19} and provide important results on rotation periods and activity cycles.  

Starting with a solar model \cite[][]{borgniet15}, we simulated complex activity patterns and the resulting brightness variations caused by spots and plages in other stars \cite[][hereafter Paper I]{meunier19}. This approach also allowed us to produce radial velocities (RV) \cite[][]{meunier19b}, astrometric variations, filling factors, and chromospheric emission \cite[see also][]{herrero16}. It is important  to understand the effect of activity on exoplanet detectability, especially in RV \cite[][]{meunier19b,meunier19c}.   
In this paper, we focus on the analysis of the complex brightness variations from the same simulations, which cover old main-sequence stars with spectral types F6-K4. These simulations are made using consistent sets of parameters, which  mostly depend on their spectral type and average activity level, for various inclinations of the star. Our objective here is to better understand how observables are related to physical parameters, which are usually either not well known or overlooked.  
Our approach is complementary to the work of \cite{shapiro14} who extended a solar model to much more active stars, which are outside of our range of parameters. However, they did not consider the effect of the spectral type nor stars that are more quiet than the Sun. 

The  paper is organized as follows: in Sect.~2 we briefly recall the main model and important parameters for the present discussion. Then we address four topics related to the diagnosis that can be made from brightness variability observations. We show the results obtained from the simulations, compare them with observations, and discuss the limitations. The topics we discuss are the following: how  brightness variability varies with spectral type (Sect.~3); whether a relationship can be established between the short-term brightness variability and other variability indicators, such as the traditional LogR'$_{\rm{HK}}$ index characterizing chromospheric emission, the long-term brightness variability, and the RV variability (in Sect.~4); whether spot sizes from brightness variability can be inferred (Sect.~5); and how to determine the meaning of the spot-dominated or plage-dominated regimes (Sect.~6). Finally, we conclude in Sect.~7.

%-----------------------------------
%-----------------------------------
\section{Models and main parameters}
%-----------------------------------
%-----------------------------------

\begin{figure}
\includegraphics*{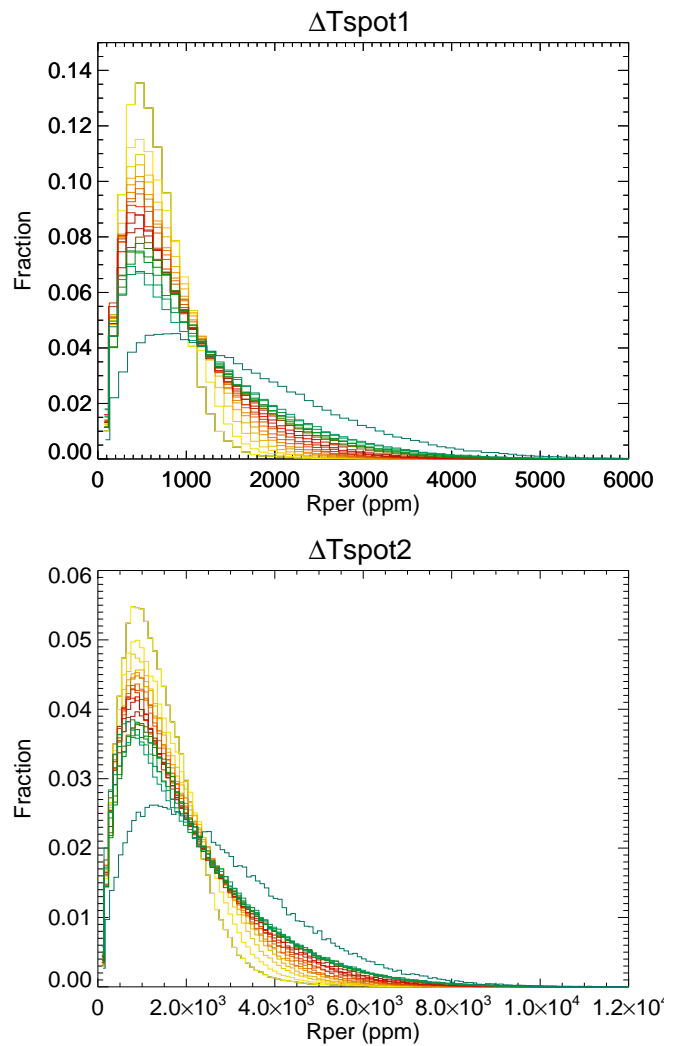}
\caption{
Normalized distributions of all individual R$_{\rm{per}}$  values for different spectral types from F6 (yellow) to K4 (blue). 
{\it Upper panel:} for $\Delta$Tspot$_1$. 
{\it Lower panel:} for $\Delta$Tspot$_2$.}
\label{rper}
\end{figure}

The model producing spots, plages, the magnetic network in a consistent way, and several observables (in particular radial velocity, photometry, and chromospheric emission) is described in detail in \cite{borgniet15} and in Paper I. At each time step, spots are injected with given properties (e.g., spatio-temporal distribution following the butterfly diagram and size distribution) after a prescribed activity cycle. A plage is then generated for each spot, assuming a certain distribution in size ratio. Both then follow a decay law and are submitted to large-scale dynamics (differential rotation and meridional circulation) and diffusion. A fraction of the remnants from the plage decay produces some network features. 
Several important parameters are adjusted to the spectral type and/or cover a range of values.\ For instance, the maximum average latitude at the beginning of the cycle $\theta_{\rm{max}}$  is not constrained for stars other than the Sun. Therefore, we study the effect of the following three values: the solar latitude $\theta_{\rm{max,\odot}}$,  $\theta_{\rm{max,\odot}}$+10$^{\circ}$, and  $\theta_{\rm{max,\odot}}$+20$^{\circ}$. Other important parameters are the rotation period P$_{\rm{rot}}$, the cycle period P$_{\rm{cyc}}$, and the cycle amplitude A$_{\rm{cyc}}$, which depend on the spectral type and on the average chromospheric activity level characterized by LogR'$_{\rm{HK}}$ .\ For each of these three parameters, we consider a median law,  a lower bound law, and an upper bound law to account for the observed dispersion among stars. Another parameter we study is the differential rotation, which  is both T$_{\rm{eff}}$-dependent and P$_{\rm{rot}}$-dependent. It also depends on the assumption made about the latitude range covered by activity.
All other parameters are kept identical to our solar values in order to limit the number of parameters. In particular, the size distributions and the distribution of the plage-to-spot size ratio are similar in all simulations.

In Paper I, the activity levels were restricted to stars with an average LogR'$_{\rm{HK}}$ below -4.5 for the most massive stars (F6) and below -4.85 for the less massive ones (K4).\ This corresponds to the plage-dominated stars of \cite{lockwood07}. The rotation rates have been deduced from activity-rotation relationships. They increase from a few days (F6 stars) to 30-70 days (K4).\ For example, G2 stars have a range of 15-32 days for G2 stars. Assuming the activity-rotation-age relationship of \cite{mamajek08}, these values correspond to ages in the range of 0.5-3 Gyrs for the most massive stars and 4-10 Gyrs for the less massive ones.

A contrast is attributed to each structure.
We used two laws for the spot temperature contrast $\Delta$Tspot: a lower bound, defined by the solar contrast \cite[as in][]{borgniet15}, and an upper bound law, depending on T$_{\rm{eff}}$ from \cite{berd05}.  We assumed real star spots have contrasts within this range. Subscripts 1 and 2 refer to these two laws. The plage contrast depends on B-V and also on the size and position of each structure.\ This has been computed for the HARPS wavelength range 378-691~nm in which a single contrast is used for the whole wavelength range, as provided by C. Norris \cite[][]{norris18}. Since our main purpose is to use these time series in conjunction with RV simulations in Paper I, the brightness variations at a one day time step were generated by summing the brightness variations caused by spots (I$_{\rm{spot}}$) and plages (I$_{\rm{plage}}$) separately (in ppm). It is important to note that I$_{\rm
plage}$  also include network features.\ Additionally, the sum of I$_{\rm{spot}}$ and I$_{\rm
plage}$ gives the total brightness variations, which is hereafter referred to as I$_{\rm{tot}}$. 

We also generated a S-index, which was then converted into a LogR'$_{\rm{HK}}$ (see Paper I for details), the filling factors covered by spots and plages, respectively, and the radial velocities. A total of 22842 time series were generated, corresponding to different sets of parameters.\ Each of these time series was produced for ten inclinations between edge-on and pole-on configurations. 

Finally, many  results presented in this paper concern the brightness short-term variability, which is routinely derived from observations in the literature. We therefore separated each time series into consecutive segments of 90 days (similar to Kepler quarters). For each segment, we eliminated points outside of the 5th -- 95th percentile range and computed the full amplitude covered by the remaining points \cite[as done in e.g.,][]{basri11}. This then gives the short-term variability R$_{\rm{per}}$. We also used the average over each time series (i.e., over all segments) of these values, $<$R$_{\rm{per}}>$.

%-----------------------------------
%-----------------------------------
\section{Effect of B-V and LogR'$_{\rm{HK}}$ on brightness variability}
%-----------------------------------
%-----------------------------------
%%%\LEt{ The first sentence of each figure legend should be a descriptive title, omitting articles (the, a, an). The text following should concisely label and explain figures and parts of figures. Please make these changes and add any additional descriptive text to the main text. Also, please avoid including references in figure legends if possible. Please do this for all figure legends.}

\begin{figure}
\includegraphics*{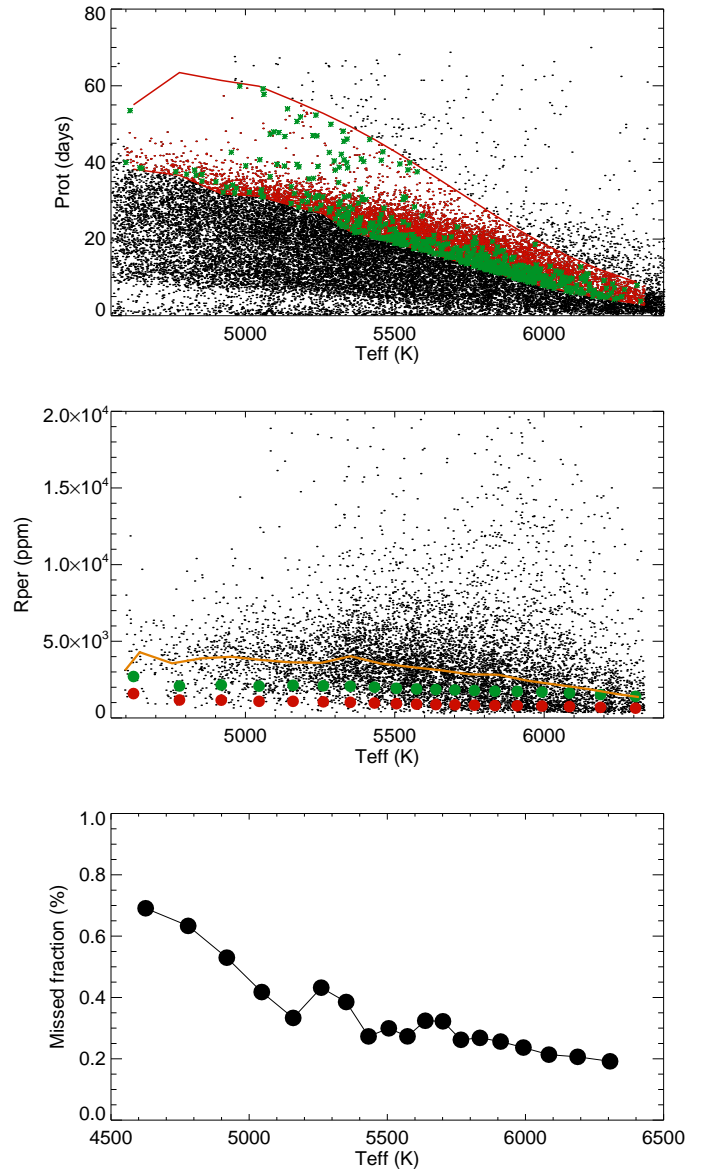}
\caption{
P$_{\rm{rot}}$ (from McQuillan et al., 2014), R$_{\rm{per}}$, and fraction of missing stars  vs. T$_{\rm{eff}}$. 
{\it  Upper panel:} The two red curves indicate our lower and upper boundaries in P$_{\rm{rot}}$. The red dots indicate stars within this P$_{\rm{rot}}$ range and green dots indicate stars  within this P$_{\rm{rot}}$ range and R$_{\rm{per}}$ above 8000 ppm.  
{\it Middle panel:} Only stars within our P$_{\rm{rot}}$  range for each spectral type are shown. The red (resp. green) points represent the median R$_{\rm{per}}$ from our simulations  vs. T$_{\rm{eff}}$, for $\Delta$Tspot$_1$ (resp. $\Delta$Tspot$_2$). The orange line is the median of the observed values within each bin.   
{\it Lower panel:} Only stars within our P$_{\rm{rot}}$  range for each spectral type are shown. 
}
\label{kepler}
\end{figure}

\begin{figure}
%\{fig_prot_rper.ps}
\includegraphics*{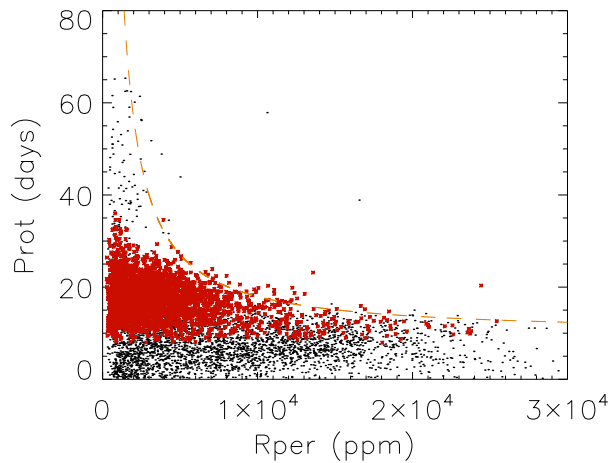}
\caption{
P$_{\rm{rot}}$ vs. R$_{\rm{per}}$ (from Mcquillan et al. 2014), for stars with T$_{\rm{eff}}$ in 5600--6000~K range. Stars with P$_{\rm{rot}}$  in the range corresponding to our simulations are highlighted in red. The orange dotted line indicates the approximate position of the upper envelope. }
\label{prot_rper}
\end{figure}

\begin{figure}
\includegraphics*{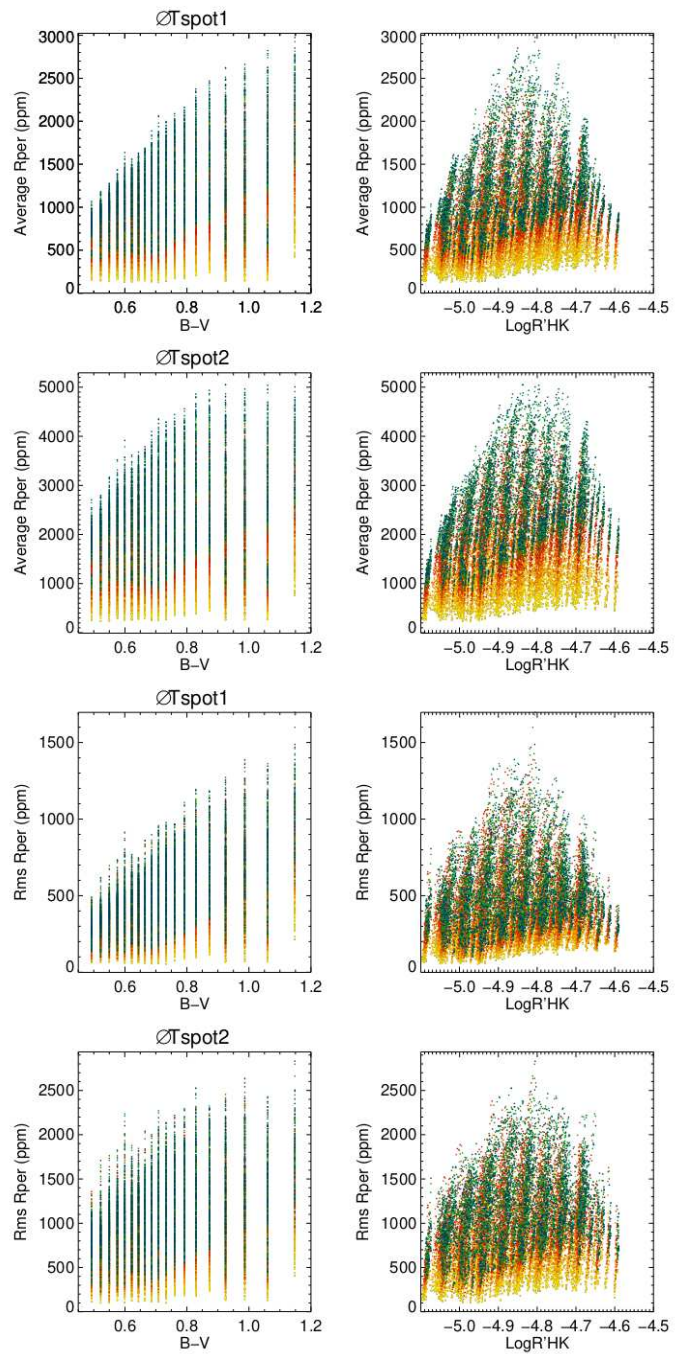}
\caption{
Average R$_{\rm{per}}$ (two first lines) and rms of individual R$_{\rm{per}}$ (two last lines) over each time series vs. B-V (left) and LogR'$_{\rm{HK}}$ (right). First and third lines correspond to $\Delta$Tspot$_1$, and second and fourth lines to $\Delta$Tspot$_2$.   
The color-coding corresponds to the inclination, from pole-on (i=0$^{\circ}$, yellow) to edge-on (i=90$^{\circ}$, blue), with light and dark orange corresponding to 20$^{\circ}$ and 30$^{\circ}$, light and dark red to 40$^{\circ}$ and 50$^{\circ}$, brown to  60$^{\circ}$, and light and dark green  to 70$^{\circ}$ and 80$^{\circ}$. Only one simulation out of five is plotted for clarity.
}
\label{rper_incl}
\end{figure}

We first address the question of the trend of the variability as a function of spectral type. In this section, we consider the averaged short-term variability defined by $<$R$_{\rm{per}}>$, as explained in Sect.~2, and the individual values on each 90 day segment. 
Figure~\ref{rper} shows the distribution of individual R$_{\rm{per}}$ for all spectral types, for which there are 19 distributions. Values of
R$_{\rm{per}}$ obtained with $\Delta$Tspot$_2$ are about twice as large as those for $\Delta$Tspot$_1$. Most values are below 4000 ppm for $\Delta$Tspot$_1$ and below 8000 ppm for $\Delta$Tspot$_2$. However, we did observe a few larger values up to $\sim$8600 and $\sim$15700, respectively. The distributions are not Gaussian and exhibit a long tail toward large R$_{\rm{per}}$. 
We observed a low increase of the brightness variability toward lower T$_{\rm{eff}}$ for both $\Delta$Tspot laws. 
The trend is most likely due to the increase in activity level on average toward K stars in our grid (i.e., corresponding to a larger number of spots and plages). This is caused by the decreasing contrast toward lower mass stars, which would produce the opposite trend if dominating. 

\subsection{Trends in the Kepler observations}

As mentioned in the Introduction (see reference there), the large number of stellar light curves obtained by Kepler have been widely  studied with the purpose of characterizing stellar activity, in particular its short-term variability (defined by R$_{\rm{per}}$) and the stellar rotation periods P$_{\rm{rot}}$. In this section, we illustrate the issues to take into account when comparing our predicted trend with the observed one. 

\subsubsection{Impact of the Kepler sampling}

The first issue is that many stars in the Kepler data set lie outside our parameter range. In fact, only a small subset lies within our P$_{\rm{rot}}$  range.\ It is important to note that this range varies with T$_{\rm{eff}}$. 
Figure~\ref{kepler} (upper panel) shows the reliable rotation periods  versus T$_{\rm{eff}}$ \cite[all data from][, Table 1 of reliable P$_{\rm{rot}}$ values]{mcquillan14}. 
Most of the stars in this table are outside our P$_{\rm{rot}}$  range and, therefore, do not correspond to our parameters as they rotate faster and are more active. In the following, 
we only selected stars within our P$_{\rm{rot}}$  range from the Mcquillan sample. In the middle panel of Fig.~\ref{kepler}, 
the resulting R$_{\rm{per}}$  versus T$_{\rm{eff}}$ dependence is shown, after selecting stars in our P$_{\rm{rot}}$  range for each spectral type. This is also from \cite{mcquillan14}.
There is a small trend toward larger R$_{\rm{per}}$ for lower T$_{\rm{eff}}$. This trend is weaker than in their original figure showing all stars.\ This occurs because there are a large number of low-mass fast-rotating stars in the global sample. For each T$_{\rm{eff}}$ bin, the P$_{\rm{rot}}$ selection impacts the distribution of R$_{\rm{per}}$ values differently. Since there   are a large number of fast-rotating stars for our lowest masses in the global sample, the average  R$_{\rm{per}}$ strongly decreases for those stars after the P$_{\rm{rot}}$ selection. We have superposed the median R$_{\rm{per}}$ derived from our simulations  versus T$_{\rm{eff}}$ for both $\Delta$Tspot
laws as a comparison. The initial result is that both trends are similar, thus the variability decreases toward higher mass stars. However, the trend is weaker for our simulations with lower R$_{\rm{per}}$ values as well. There are also a few Kepler light curves that provide much higher R$_{\rm{per}}$ than in our models (see Sect.~3.3).

\subsubsection{Impact of biases in P$_{\rm{rot}}$  determination}

Another   issue that makes this comparison more complicated is that there are very few stars with P$_{\rm{rot}}$  above 40 days from observations (as seen
in the upper panel). This leads to many missing points corresponding to low activity at first glance.  Even though we modeled these stars, we expected a low R$_{\rm{per}}$. In addition to the table of reliable P$_{\rm{rot}}$ used above, \cite{mcquillan14} also provide a larger table (their table 2) with stars with unreliable P$_{\rm{rot}}$. We used both tables to estimate the percentage of missed stars in the sample of reliable P$_{\rm{rot}}$. This bias in observation is shown on the lower panel. Up to 70\% of the low-mass stars are missing for our P$_{\rm{rot}}$  range. This  would be the opposite when considering all stars. Therefore, the observed trend is strongly biased. This may explain part of the difference we observed since poorly determined P$_{\rm{rot}}$  could be due to low activity stars with a low R$_{\rm{per}}$, especially for low-mass stars. 
We note that the bias estimation itself has some uncertainties since unreliable P$_{\rm{rot}}$  values are used to compute it. 

Finally, we computed the amplitude of the autocorrelation
function at P$_{\rm{rot}}$  ,which is known in our simulations. However, it is important to note that it was beyond the scope of our paper to fully explore the temporal variability of the simulated time series and, in particular, the rotation rate.%for example as done by \cite{mcquillan14}. 
We found that  the autocorrelation takes large values for the stars with the largest T$_{\rm{eff}}$ in our grid. However, for late G and K stars, the autocorrelation function at P$_{\rm{rot}}$  drops and even some negative values can be observed.\ This shows that P$_{\rm{rot}}$  is indeed difficult or even impossible  to measure for some of the stars in our activity range, at least on short time series such as the Kepler quarters. One reason for this is that as P$_{\rm{rot}}$  increases, the lifetime of the structures becomes low compared to P$_{\rm{rot}}$  and the coherence is lost. This is consistent with the bias discussed above. 

These difficulties also impact important properties of the P$_{\rm{rot}}$ distributions, 
such as the dearth of intermediate rotation periods \cite[][]{mcquillan13b,davenport17}. This was discussed by \cite{reinhold19}, who showed that 
configurations, in which there is equilibrium between spots and plages, can lead to very unreliable or
impossible to measure P$_{\rm{rot}}$.\ This may explain the dearth.

\subsubsection{Impact of parameter distribution in the simulations}

Finally, a third issue is that the distributions we have shown in Fig.~\ref{rper} for our simulated R$_{\rm{per}}$ correspond to all simulations in the grid. Each simulation counts as one across the grid of parameters. However, some values may be more likely to appear than others for some parameters. Additionally, the distribution in cycle amplitude is not necessarily homogeneous and the distribution in $\Delta$Tspot  versus T$_{\rm{eff}}$ is unknown. Even though the trend obtained from simulations should be reliable since they show how parameters impact the variability, parameter distributions may also impact this trend as well.

\subsection{Stars with a large R$_{\rm{per}}$}

\begin{figure}
\includegraphics*{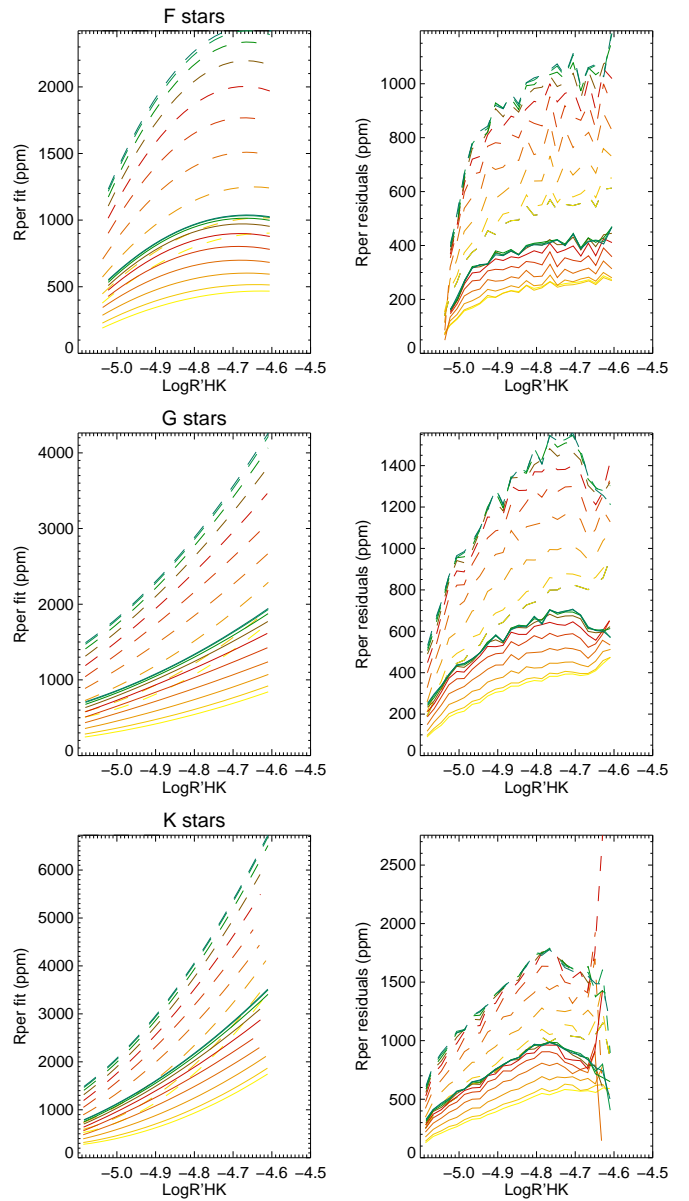}
\caption{
Polynomial fits to R$_{\rm{per}}$  vs. LogR'$_{\rm{HK}}$ (left panels) for different inclinations and residuals (binned in LogR'$_{\rm{HK}}$) after subtraction of polynomial fit (left panels), for $\Delta$Tspot$_1$ (resp. $\Delta$Tspot$_2$) shown as solid (resp. dashed) lines. F, G, and K stars are shown separately. The color-coding corresponds to the inclination, from pole-on (i=0$^{\circ}$, yellow) to edge-on (i=90$^{\circ}$, blue), with light and dark orange corresponding to 20$^{\circ}$ and 30$^{\circ}$, light and dark red to 40$^{\circ}$ and 50$^{\circ}$, brown to  60$^{\circ}$, and light and dark green  to 70$^{\circ}$ and 80$^{\circ}$.  
}
\label{rper_logrphk}
\end{figure}

There are a few stars with very large R$_{\rm{per}}$ in the observations that are still within our P$_{\rm{rot}}$ range. This is difficult to reproduce in simulations. These are shown in green in the upper panel of Fig.~\ref{kepler}. We find that they can be classified in two categories of points.  Firstly, points at relatively large masses (T$_{\rm{eff}}>$5500K) are very close to the lower bound in P$_{\rm{rot}}$ and represent a very small number of stars (about 1-2\%).
Secondly, points at relatively low masses (T$_{\rm{eff}}<$5500K) are spread over the whole range in P$_{\rm{rot}}$ and represent  about 7\% more stars. They are close to the lower boundary in P$_{\rm{rot}}$ and  rotate relatively quickly. An analysis of P$_{\rm{rot}}$   versus R$_{\rm{per}}$ from the data of \cite{mcquillan14} shown in Fig.~\ref{prot_rper},  shows that the upper envelope decreases. Faster rotation leads to larger R$_{\rm{per}}$; nevertheless, low R$_{\rm{per}}$  are also possible for fast rotators. It is important to note that this envelope is quite flat for large R$_{\rm{per}}$. The points corresponding to the lowest P$_{\rm{rot}}$  in our range lie there. Consequently, even a slight uncertainty in P$_{\rm{rot}}$  would therefore easily put these stars outside of our R$_{\rm{per}}$  range. 
It is not surprising to observe a few more active stars within our  range of P$_{\rm{rot}}$  with a larger R$_{\rm{per}}$ than what is simulated.  

 When considering this particular selection of points with high activity level, we observed a trend in which   P$_{\rm{rot}}$ increased for the low-mass stars in our grid. This means that for a similar activity level, as defined by R$_{\rm{per}}$, low-mass stars have a larger P$_{\rm{rot}}$. This trend is consistent with the activity-rotation relationship in the literature \cite[e.g.,][]{noyes84b,saar99,mamajek08}, which was also used to build our synthetic time series (Sect.~2).  
In conclusion, the variability derived from our simulation exhibits similarities with observations.\ Nevertheless, comparisons must be made with caution since possible biases due to the sampling and parameter distributions can occur.

\section{Relationship between R$_{\rm{per}}$ and other activity}

In this section, we establish relationships between R$_{\rm{per}}$ and other activity indicators. We first consider its relation with the average and the rms LogR'$_{\rm{HK}}$. We then consider its relation with the long-term brightness variability and the RV variations. 

\subsection{Relationship between R$_{\rm{per}}$ and average LogR'$_{\rm{HK}}$}

Figure~\ref{rper_incl} shows R$_{\rm{per}}$  versus B-V, which illustrates the increase with rising B-V (Sect.\ 3), versus the average LogR'$_{\rm{HK}}$. The average R$_{\rm{per}}$ shown on  the first two lines of the figure, is strongly impacted by inclination. The short-term variability level is lower for pole-on configurations than for edge-on configurations by a factor of about two to three. This was already seen for the Sun \cite[][]{borgniet15}. This is caused by the rotational modulation which is much lower when inclination decreases. There is still a significant variability for pole-on configurations, which is probably due to the limited lifetime of the structures. On the other hand, the dispersion in R$_{\rm{per}}$ which is defined as the rms of the array of the individual 90-day measurements, and computed for each time series. Each of these rms is representative of a long-term variability along the cycle. These rms do not depend as much on inclination.\ However, there is a definite trend, as seen before for the Sun \cite[][]{knaack01,borgniet15}. 
The standard deviation (hereafter rms) of R$_{\rm{per}}$ increases with B-V. The ratio between this dispersion and the average is relatively flat with respect to B-V and LogR'$_{\rm{HK}}$.

Many results on stellar activity over the past decades have been published based on LogR'$_{\rm{HK}}$ \cite[e.g.,][]{noyes84,baliunas95,radick98,gray03,gray06,lockwood07,hall09,mittag13,hempelmann16,mittag17,radick18}. It is a very widely used activity indicator to derive empirical laws \cite[e.g.,][to study the age-activity-rotation relation]{mamajek08}.
However, over the last few years, an increasing number of results have been obtained using photometric data, in particular Kepler, using the short-term variability indicator R$_{\rm{per}}$ (or equivalent measurements). These studies also lead to empirical relationships between R$_{\rm{per}}$ and other variables (e.g., P$_{\rm{rot}}$ and differential rotation). 
Therefore, it would be useful to clarify the relationship between the two for future reference. 
The two indicators do not trace the same information. The LogR'$_{\rm{HK}}$ indicator represents the emission in plage and network with no degeneracy with spots, while R$_{\rm{per}}$ is related to both spots and plages and has a strong degeneracy between the contributions of these two types of structures. 

The relationship between R$_{\rm{per}}$ and LogR'$_{\rm{HK}}$ shown in Fig.~\ref{rper_incl} has a very large dispersion due to inclination. In addition, R$_{\rm{per}}$ is also very sensitive to $\Delta$Tspot, which is not the case for LogR'$_{\rm{HK}}$. For each inclination and $\Delta$Tspot, we have fit R$_{\rm{per}}$   versus LogR'$_{\rm{HK}}$ using a polynomial function. The result is shown in Fig.~\ref{rper_logrphk} for F, G, and K stars, respectively. The possible range in $\Delta$Tspot values we have considered typically introduces a factor of two on the variability. The full range of inclinations leads to a similar factor. The figure also shows the rms of the residuals after removing these fits, which cover a wide range. There is much overlap between different inclinations and $\Delta$Tspot. For example, for G stars and LogR'$_{\rm{HK}}\sim$-4.8, high inclination R$_{\rm{per}}$ and $\Delta$Tspot$_1$ overlap with low inclination R$_{\rm{per}}$ and $\Delta$Tspot$_2$. 

We conclude that although R$_{\rm{per}}$ and LogR'$_{\rm{HK}}$ are related, their relative amplitude strongly varies, at least with spectral type, spot temperature, and inclination.\ Additionally, the correlation between the two indicators is poor. Therefore, it is important to be very careful when comparing laws from various sources, unless these factors can be taken into account. It may also be difficult to separate low inclination higher activity stars and high inclination, low activity stars from R$_{\rm{per}}$ diagnosis.  

\subsection{Relationship between R$_{\rm{per}}$ and LogR'$_{\rm{HK}}$ variability}

\begin{figure}
\includegraphics*{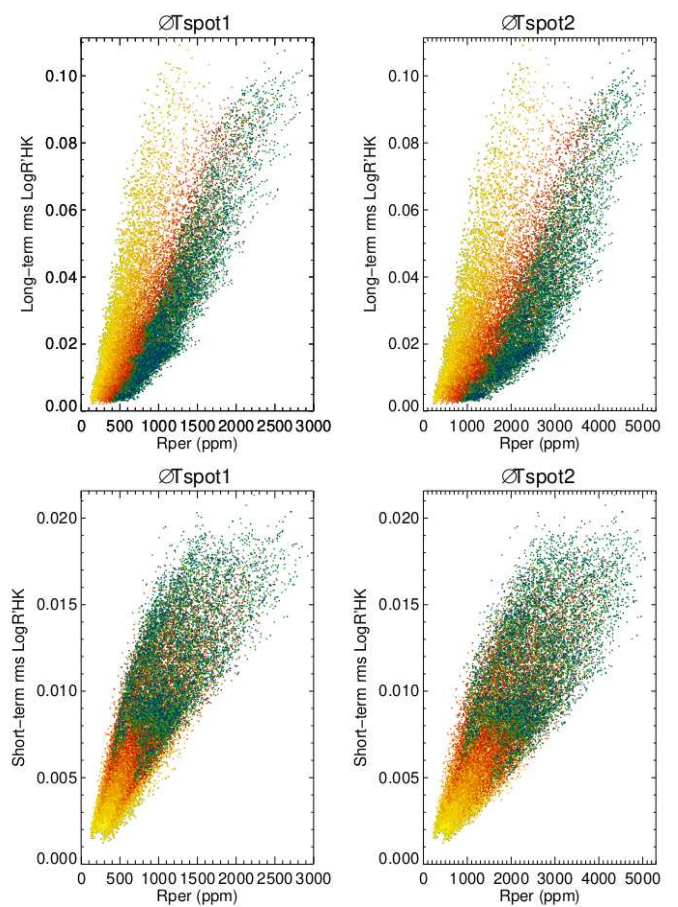}
\caption{
Rms of long-term variability of LogR'$_{\rm{HK}}$  (upper panel) and of short-term variability (lower panel) vs. average R$_{\rm{per}}$ for each time series, for $\Delta$Tspot$_2$ (left) and $\Delta$Tspot$_2$ (right). The color-coding corresponds to the inclination, from pole-on (i=0$^{\circ}$, yellow) to edge-on (i=90$^{\circ}$, blue), with light and dark orange corresponding to 20$^{\circ}$ and 30$^{\circ}$, light and dark red to 40$^{\circ}$ and 50$^{\circ}$, brown to  60$^{\circ}$, and light and dark green  to 70$^{\circ}$ and 80$^{\circ}$. 
Only one simulation out of five is plotted for clarity. 
}
\label{rper_rmslogrphk}
\end{figure}

On the other hand, Fig.~\ref{rper_rmslogrphk} shows that R$_{\rm{per}}$ is slightly better related to the  LogR'$_{\rm{HK}}$ variability as compared to the average activity level.  This is the case even though R$_{\rm{per}}$ is a residual between spot and plage contributions, contrary to LogR'$_{\rm{HK}}$. The upper panels show that for a given inclination and $\Delta$Tspot, there is a reasonable correlation between the two. The short-term variability in LogR'$_{\rm{HK}}$ (lower panels) is also correlated with the average R$_{\rm{per}}$. The inclination effect then dominates the relationship. In that case, it is mostly due to both short-term variabilities, in R$_{\rm{per}}$ and LogR'$_{\rm{HK}}$, that increase when going from pole-on to edge-on.\ This is due to the increase in the rotation-modulated signal. 

\subsection{Relationship between R$_{\rm{per}}$ and long-term brightness variability}

\begin{figure}
\includegraphics*{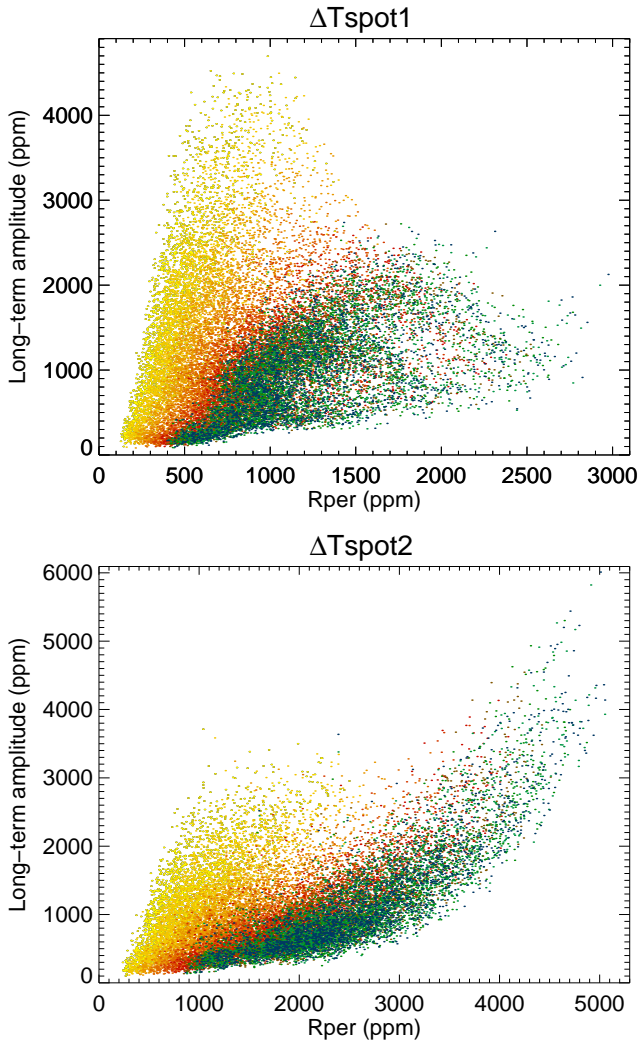}
\caption{
Long-term amplitude of variability (in absolute value)  vs. average R$_{\rm{per}}$ for each time series for $\Delta$Tspot$_1$ (upper panel) and $\Delta$Tspot$_2$ (lower panel). The color-coding corresponds to the inclination, from pole-on (i=0$^{\circ}$, yellow) to edge-on (i=90$^{\circ}$, blue), with light and dark orange corresponding to 20$^{\circ}$ and 30$^{\circ}$, light and dark red to 40$^{\circ}$ and 50$^{\circ}$, brown to  60$^{\circ}$, and light and dark green  to 70$^{\circ}$ and 80$^{\circ}$. 
 Only one simulation out of five is plotted for clarity.
}
\label{rperlt}
\end{figure}

Although the short-term variability is easily characterized from Kepler data, for example, this is not the case for long-term variability. Here, we investigate how the short-term variability is related to the long-term variability as defined by the difference between maximum and minimum on each one year smoothed time series ($\Delta$I hereafter). 
Figure~\ref{rperlt} shows the absolute value of $\Delta$I  versus the average R$_{\rm{per}}$. There is a large dispersion, which is mostly due to inclination but also $\Delta$Tspot. 
For lower inclinations (yellow), the long-term amplitude is much larger compared to edge-on configuration for a similar R$_{\rm{per}}$ level. The short term variability is twice as large for $\Delta$Tspot$_2$ compared to $\Delta$Tspot$_1$ (as seen in Sect.~3).
On the other hand, the long-term variability is lower for $\Delta$Tspot$_2$ at low inclinations and larger for $\Delta$Tspot$_2$ at large inclinations.
The two plots therefore differ qualitatively and the relation between short-term and long-term variability is complex for the type of stars we consider here.

\begin{figure}
\includegraphics*{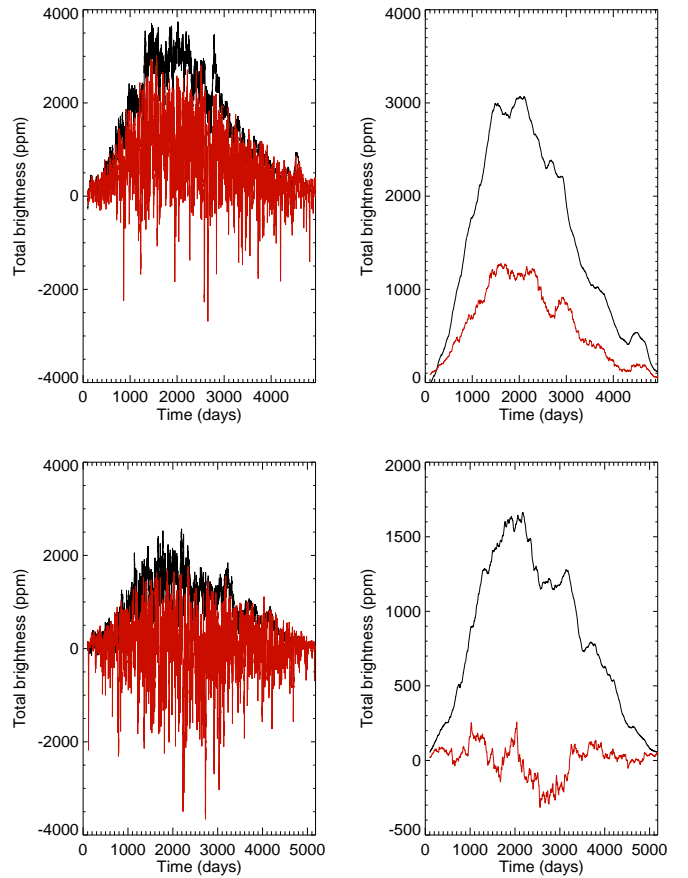}
\caption{
Brightness variation for star with similar long-term variability (upper panel) when seen pole-on (black) and seen edge-on (red), and with reversal between inclinations (lower panel), with no smoothing (left) and with smoothing (right).  
}
\label{DI}
\end{figure}
%ratio 2.56 en haut et -2.83 en bas
%print,prot(num_simu),pcyc(num_simu),bv(num_simu),typesp(num_simu),logrphk_sel(num_simu),latitude(num_simu),m(num_simu)
%38.653000       13.541000      0.82880000G9      _4.9569000       20.500000       2.0000000
%51.780000       14.190000       1.0607000K3      -4.9914000       20.500000       2.0000000

\begin{table}
\caption{Long-term amplitude $\Delta$I for pole-on and edge-on configurations}
\label{tabDI}
\begin{center}
\renewcommand{\footnoterule}{}  % to avoid a line before footnotes
\begin{tabular}{lll}
\hline
&  $\Delta$Tspot$_1$ & $\Delta$Tspot$_2$ \\
 \hline
Dominant feature &  $\Delta$I$_{0}>$0 & $\Delta$I$_{90}<$0 \\
   &   99\%   &   97\% \\
 \hline
Case $\Delta$I$_{0}$,$\Delta$I$_{90}$ same sign & all $>$0    & all$<$0    \\
Fraction & 90\% & 18\% \\
Median $\Delta$I$_{0}$/$\Delta$I$_{90}$ & 1.82 & 0.97 \\
Range $\Delta$I$_{0}$/$\Delta$I$_{90}$ & [1.35;2.56] & [0.40;2.18] \\
\hline
Case $\Delta$I$_{0}$,$\Delta$I$_{90}$ opposite sign & $\Delta$I$_{0}>$0    & $\Delta$I$_{0}>$0    \\
 & $\Delta$I$_{90}<$0    & $\Delta$I$_{90}<$0    \\
Fraction & 10\% & 78\% \\
Median $\Delta$I$_{0}$/$\Delta$I$_{90}$ & -1.42 & -1.23 \\
Range $\Delta$I$_{0}$/$\Delta$I$_{90}$ & [-2.83;-0.64] & [-2.64;-0.44] \\
\hline
\end{tabular}
\end{center}
\tablefoot{Percentages are given with respect to the total sample. The ranges correspond to the fifth and the ninth percentiles over all $\Delta$I$_{0}$ and $\Delta$I$_{90}$ values}.
\end{table}

We now study, in more detail, the sign and amplitude of the long-term variability across our stellar grid for stars seen pole-on (i=0$^{\circ}$) and edge-on (i=90$^{\circ}$)\footnote{Subscripts 0 and 90  to  $\Delta$I correspond to the pole-on and edge-on configurations respectively}. 
A summary of the different configurations is shown in Table~\ref{tabDI}. 
For $\Delta$Tspot$_1$, $\Delta$I$_{0}$ is positive most of the time (i.e., larger flux at cycle maximum). In 90\% of the cases, $\Delta$I$_{90}$ has the same sign and both are then plage-dominated (see Sect.~6). This is in agreement with the results of \cite{shapiro14} for a solar-like activity pattern (i.e., an increase in variability at low inclination). For the remaining 10\%, however, $\Delta$I$_{90}$ is negative and leads to a reversal between the two.
%%% The ratio covers a large range of values, and shows a weak trend with B-V. 
For $\Delta$Tspot$_2$, $\Delta$I$_{90}$ is usually negative while $\Delta$I$_{0}$ can be both negative or positive, or shift more toward spot-dominated configurations. 
%Again, the ratio is on average different from 1 (in absolute value) and can be either lower or larger than one.  
An illustration of two extreme cases is shown in Fig.~\ref{DI}.

In conclusion, we observe large differences in amplitude between pole-on and edge-on brightness amplitudes.\  Oftentimes, there is a reversal in sign between the two inclinations. We will study the spot and plage regimes in more detail in Sect.~6.  

\subsection{Relationship between R$_{\rm{per}}$ and RV variability}

\begin{figure}
\includegraphics*{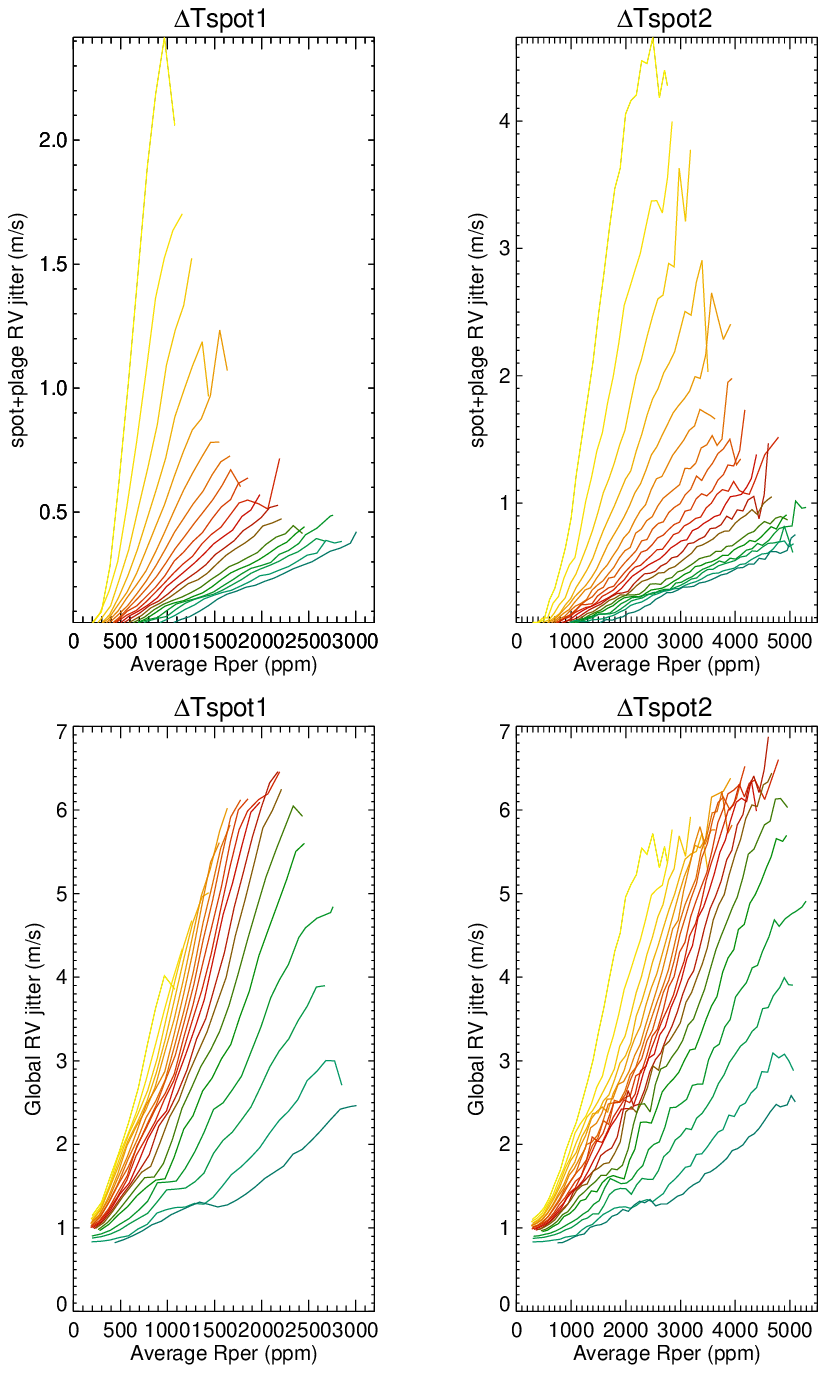}
\caption{
RV jitter vs. R$_{\rm{per}}$ (binned), for 19 spectral types (color code similar to Fig.~1), for $\Delta$Tspot$_1$ (left) and $\Delta$Tspot$_2$ (right). {\it Upper panels}: RV jitter due to spots and plages.  
{\it Lower panels}: Full RV jitter (activity, oscillation+granulation+supergranulation averaged over six hours, 0.6 m/s white noise). 
}
\label{rper_rv}
\end{figure}

\begin{table*}
\caption{RV jitter  vs. R$_{\rm{per}}$ linear fit}
\label{sloperv}
\begin{center}
\renewcommand{\footnoterule}{}  % to avoid a line before footnotes
\begin{tabular}{lllllllll}
\hline
Spectral & offset  & slope  & offset & slope & offset  & slope  & offset & slope \\ 
Type &  (all) & (all)  & (all) & (all)  & (90$^{\circ}$) & (90$^{\circ}$) & (90$^{\circ}$) & (90$^{\circ}$) \\ 
   & $\Delta$Tspot$_1$ & $\Delta$Tspot$_1$ & $\Delta$Tspot$_2$ & $\Delta$Tspot$_2$ & $\Delta$Tspot$_1$ & $\Delta$Tspot$_1$ & $\Delta$Tspot$_2$ & $\Delta$Tspot$_2$ \\
\hline
 F6  &   0.3  &   3.6e-03  &   0.2  &   2.2e-03  &   -0.2  &   4.3e-03  &   -0.4  &   2.5e-03 \\
 F7  &   0.2  &   3.4e-03  &   0.2  &   1.9e-03  &   -0.8  &   4.6e-03  &   -0.8  &   2.3e-03 \\
 F8  &   0.2  &   3.4e-03  &   0.1  &   1.7e-03  &   -1.4  &   4.9e-03  &   -1.3  &   2.3e-03 \\
 F9  &   0.1  &   3.4e-03  &   0.2  &   1.6e-03  &   -1.5  &   4.8e-03  &   -1.2  &   2.1e-03 \\
 G0  &  -0.0  &   3.6e-03  &   0.1  &   1.6e-03  &   -1.3  &   4.5e-03  &   -1.2  &   2.0e-03 \\
 G1  &  -0.0  &   3.4e-03  &   0.1  &   1.5e-03  &   -1.7  &   4.7e-03  &   -1.3  &   2.0e-03 \\
 G2  &  -0.1  &   3.3e-03  &   0.0  &   1.5e-03  &   -1.6  &   4.4e-03  &   -1.4  &   2.0e-03 \\
 G3  &  -0.1  &   3.4e-03  &  -0.0  &   1.5e-03  &   -1.9  &   4.5e-03  &   -1.7  &   2.0e-03 \\
 G4  &  -0.1  &   3.2e-03  &  -0.1  &   1.5e-03  &   -1.7  &   4.2e-03  &   -1.6  &   1.9e-03 \\
 G5  &  -0.1  &   3.1e-03  &  -0.0  &   1.4e-03  &   -1.4  &   3.8e-03  &   -1.4  &   1.8e-03 \\
 G6  &  -0.2  &   3.0e-03  &  -0.1  &   1.4e-03  &   -1.9  &   4.0e-03  &   -1.7  &   1.9e-03 \\
 G7  &  -0.3  &   3.0e-03  &  -0.2  &   1.4e-03  &   -1.9  &   3.8e-03  &   -1.8  &   1.8e-03 \\
 G8  &  -0.2  &   2.8e-03  &  -0.2  &   1.4e-03  &   -2.2  &   3.8e-03  &   -2.0  &   1.8e-03 \\
 G9  &  -0.3  &   2.5e-03  &  -0.2  &   1.3e-03  &   -2.1  &   3.4e-03  &   -2.0  &   1.7e-03 \\
 K0  &  -0.2  &   2.2e-03  &  -0.2  &   1.1e-03  &   -1.8  &   3.0e-03  &   -1.7  &   1.5e-03 \\
 K1  &   0.0  &   1.7e-03  &   0.1  &   8.9e-04  &   -0.9  &   2.1e-03  &   -0.9  &   1.1e-03 \\
 K2  &   0.2  &   1.3e-03  &   0.2  &   7.0e-04  &   -0.5  &   1.6e-03  &   -0.5  &   8.4e-04 \\
 K3  &   0.4  &   8.7e-04  &   0.3  &   5.0e-04  &   -0.2  &   1.1e-03  &   -0.2  &   6.3e-04 \\
 K4  &   0.4  &   6.3e-04  &   0.4  &   4.0e-04  &   -0.6  &   1.0e-03  &   -0.6  &   6.1e-04 \\
\hline
\end{tabular}
\end{center}
\tablefoot{The offsets and slopes are given for all inclinations (averaged, corresponding to the plots of Fig.~7) or only for edge-on configurations (90$^{\circ}$), for both spot contrasts. }
\end{table*}

\begin{figure}
\includegraphics*{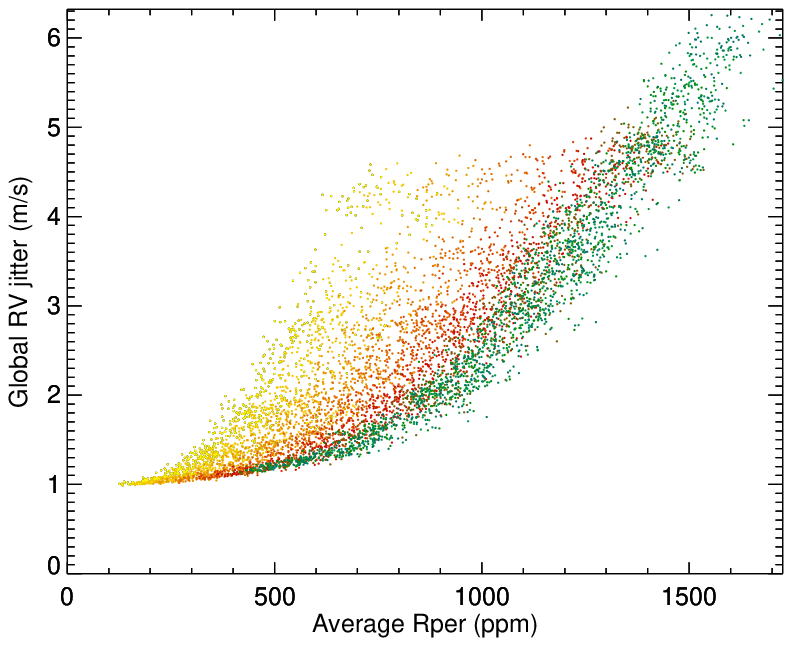}
\caption{
RV jitter  vs. R$_{\rm{per}}$ for all G2 simulations, and different inclinations, for $\Delta$Tspot$_1$. The color-coding corresponds to the inclination, from pole-on (i=0$^{\circ}$, yellow) to edge-on (i=90$^{\circ}$, blue), with light and dark orange corresponding to 20$^{\circ}$ and 30$^{\circ}$, light and dark red to 40$^{\circ}$ and 50$^{\circ}$, brown to  60$^{\circ}$, and light and dark green  to 70$^{\circ}$ and 80$^{\circ}$. 
}
\label{rper_rv_G2}
\end{figure}

Another useful relationship is the relation between R$_{\rm{per}}$ and the RV variability. Indeed, it would be interesting to be able to predict the RV variability from the brightness variability (e.g., in order to select the best targets for Kepler, TESS, or PLATO follow-ups).  
%\textbf{ voir ce qu'on cite; \cite{aigrain12} derivation de RV d'apres methode ff' ; mais cinomplet ? tester cette methode ??  + RULE OF THUMB THESE HAYWOOD : mais n'etait donne qu'un ordre de grandeur, voir si citer ou pas + DISCUTER / ou pas de citation du tout ici}. 
Figure~\ref{rper_rv} shows the average relation between these variables individually for the 19 spectral types. We first considered the relationship with RV solely because of the spot and plage contrasts (upper panel of Fig.~\ref{rper_rv}). Since it is superposed to other contributions, this is not an observable.\ However, it is useful for a better understanding as it is the RV component directly related to the photometric variability. We find that for a given spectral type the laws are relatively linear, but the slope strongly depends on spectral type. 
This is because contrast only plays a role for the brightness variations  (brightness variations also depend on the size distribution and spatio-temporal distribution as the RV, which are the same for all spectral types here), while the RV also depends on P$_{\rm{rot}}$  for example. This leads to a much lower slope for K4 stars compared to F6 stars. 

The lower panels show the same plots. However, the RV jitter corresponding to the total RV signal have  a binning over each spectral type
and over inclinations. The RV signal includes spot+plage, convective blueshift inhibition, oscillation, granulation, and supergranulation. The last three are  averaged over one hour and a 0.6 m/s instrumental noise. Again, there is a strong spectral type effect.  The slopes and offsets are presented in Table~\ref{sloperv} after averaging all inclinations (first columns). 
Inclination adds a systematic shift to these curves, as illustrated for G2 stars in Fig.~\ref{rper_rv_G2}.\ Here, the slopes are not very different, but there is a clear offset between inclinations. The factor between the most extreme inclinations is in the range 1.5-2.  
Table~\ref{sloperv} also shows the slope and offset for the edge-on configuration only.  
There is a strong spot temperature effect as well, because R$_{\rm{per}}$ is very sensitive to this parameter.\ It is important to note that there are lower slopes when the contrast is larger. However, the RV variations, being the superposition of several components, are such that $\Delta$Tspot does not have a major effect on RV. As a consequence the slopes strongly depend on the spot contrast. Finally, the relationships are not entirely linear, as low R$_{\rm{per}}$ values have a lower slope. 
Nevertheless, there is a clear relationship between the short-term photometric
variability R$_{\rm{per}}$ and the RV jitter for very specific conditions, such as spectral type, inclination, spot contrast, and P$_{\rm{rot}}$. These relationships strongly depend on the type of stars and its properties, and cover a large range of slope values.

%-----------------------------------
%-----------------------------------
\section{Spot sizes from brightness variability}
%-----------------------------------
%-----------------------------------

\begin{figure}
\includegraphics*{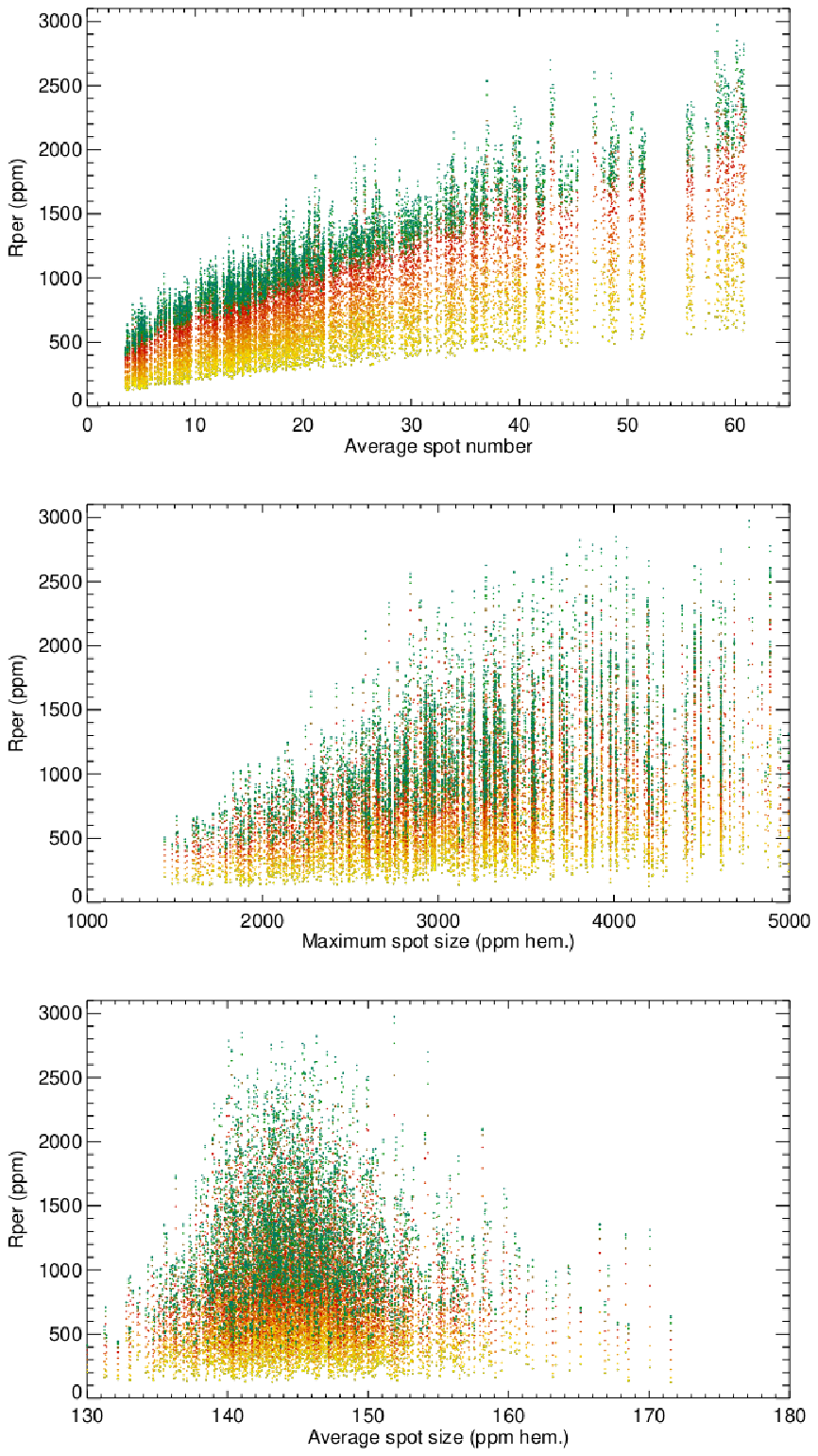}
\caption{
R$_{\rm{per}}$  vs. average (over time) spot number (upper panel), vs. maximum spot size (middle panel), and  vs. average spot size  (lower panel). 
The color-coding corresponds to the inclination, from pole-on (i=0$^{\circ}$, yellow) to edge-on (i=90$^{\circ}$, blue), with light and dark orange corresponding to 20$^{\circ}$ and 30$^{\circ}$, light and dark red to 40$^{\circ}$ and 50$^{\circ}$, brown to  60$^{\circ}$, and light and dark green  to 70$^{\circ}$ and 80$^{\circ}$. 
Only one simulation out of five is plotted for clarity.
}
\label{spotsize}
\end{figure}

The short-term brightness variability has been interpreted as an indication of star spot size \cite[][]{giles17}, from which an empirical law of spot size  versus T$_{\rm{eff}}$ was derived. The argument was the following: assuming a large spot size distribution for a given star, as for the Sun, there must be a dominating spot, and R$_{\rm{per}}$ must then be representative of that particular spot size. 
This analysis does not take into account  plages that also play a role, it is well known that R$_{\rm{per}}$ is only a residual from the spot and plage contributions (see Sect. 6).\ Furthermore,   sizes are strongly degenerated with the contrast. In addition, we show that the assumptions made are not valid here, at least for the stars within our range of activity levels (i.e., relatively low activity).

In our simulations, we always use the same solar spot size distribution. Therefore, the average number of spots  for each simulation is fairly constant (mostly in the range 130--160 ppm), as shown in the lower panel of Fig.~\ref{spotsize}.\ This does not prevent R$_{\rm{per}}$  from being highly variable. The random generation of spots for a large amount of realizations also shows that dominating spots exist, but they are rare.\ In our case, there are less than a few percent of dominating spots on the assumption that they should represent more than half of the whole spot filling factor. 

In addition, R$_{\rm{per}}$ is strongly dependent on the spot number. In fact, it is the spot number that controls the variability and not the spot size here. It is also possible to obtain time series with a large maximum spot size but a low R$_{\rm{per}}$ (middle panel).\ This is because there is a large dispersion in this relation in addition to the inclination effect.

Therefore, it is important not to overestimate the importance of information from brightness variability. It is without a doubt also related to the spot size distribution in some way.\ However, there is a strong degeneracy with many other important properties, which are not constrained either. This is probably true as well for lifetime estimation using auto-correlations, although to a lesser extent. They are likely indicative of an active region lifetime altogether, but this is not based on individual features.

%-----------------------------------
%-----------------------------------
\section{Plage-dominated and spot-dominated regimes}
%-----------------------------------
%-----------------------------------

\begin{figure}
\includegraphics*{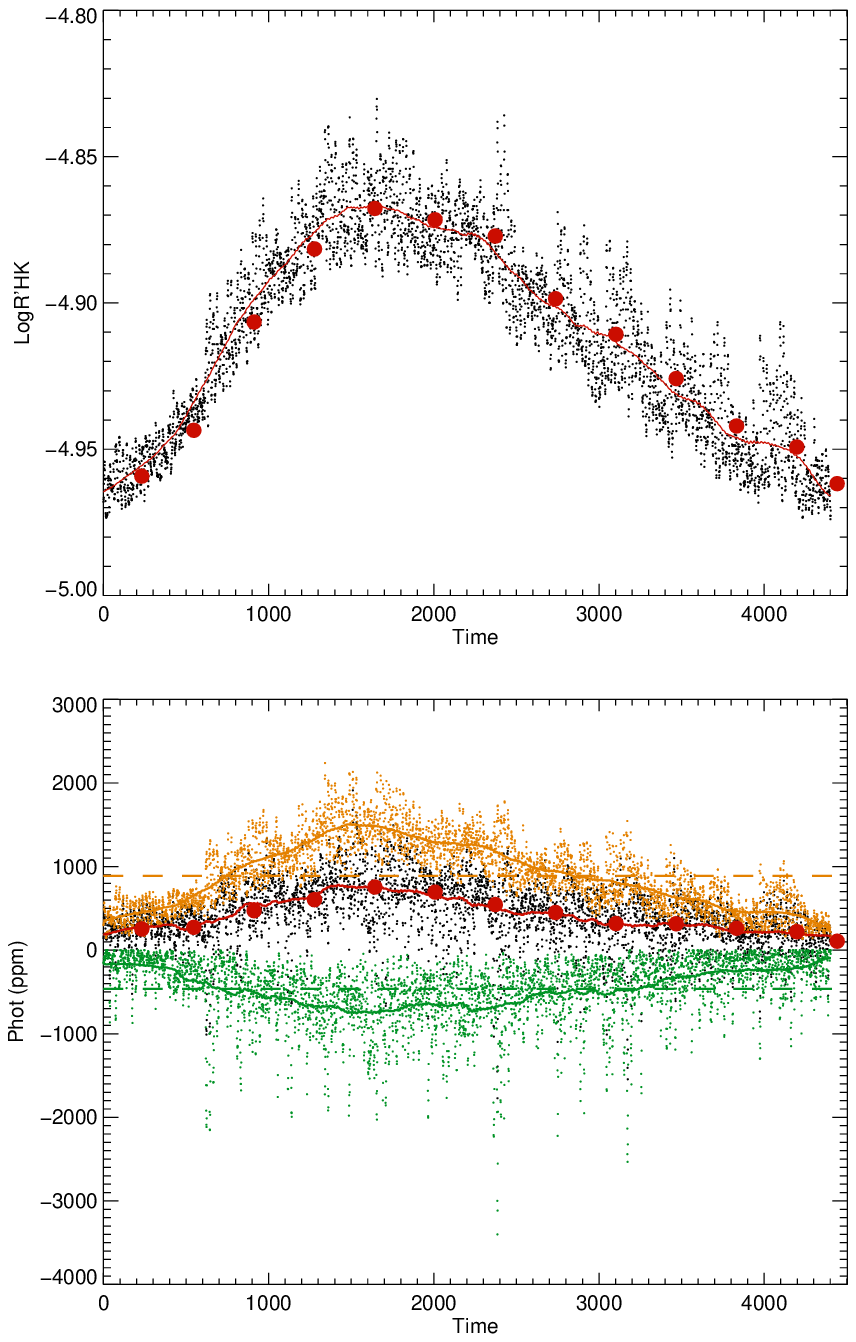}
\caption{
LogR'$_{\rm{HK}}$  (upper panel) and I$_{\rm{tot}}$ (lower panel)  vs. time.
{\it Upper panel:} 
LogR'$_{\rm{HK}}$  vs. time for a moderately active G2 star, seen edge-on (black dots). The red curves is the smoothed (over one year) series, and the red points are yearly averages. 
{\it Lower panel:} 
Same for I$_{\rm{tot}}$  vs. time. The orange dots only represent the plage contribution (the dashed orange line indicates the average level), and the green dots represent the spot contribution (the horizontal dashed green lines indicates the average level) for $\Delta$Tspot$_1$.
}
\label{ica_typique}
\end{figure}

\begin{figure}
\includegraphics*{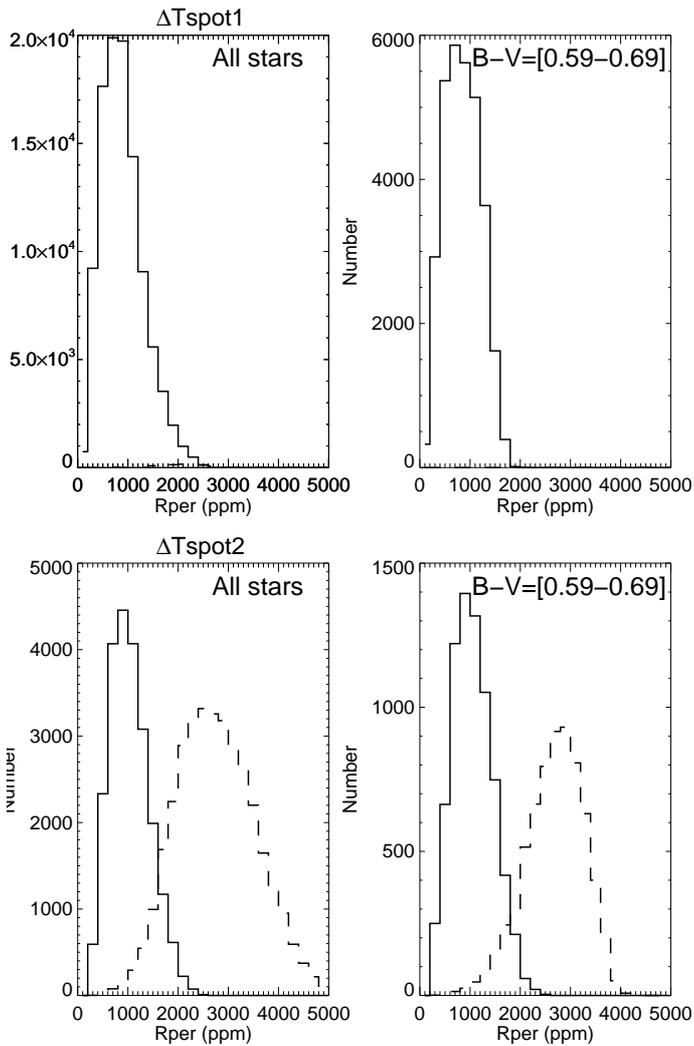}
\caption{
Distribution of number of simulations with correlation $C$ between LogR'$_{\rm{HK}}$ and I$_{\rm{tot}}$ larger than 0.6 (solid line) and lower than -0.6 (dashed line) for  $\Delta$Tspot$_1$ (upper panels) and $\Delta$Tspot$_2$ (lower panels). 
The left panels are for all spectral types and the right panels  for stars with B-V between 0.59 and 0.69 (for a comparison with Montet et al. 2017).
}
\label{dist_correl}
\end{figure}

\begin{figure}
\includegraphics*{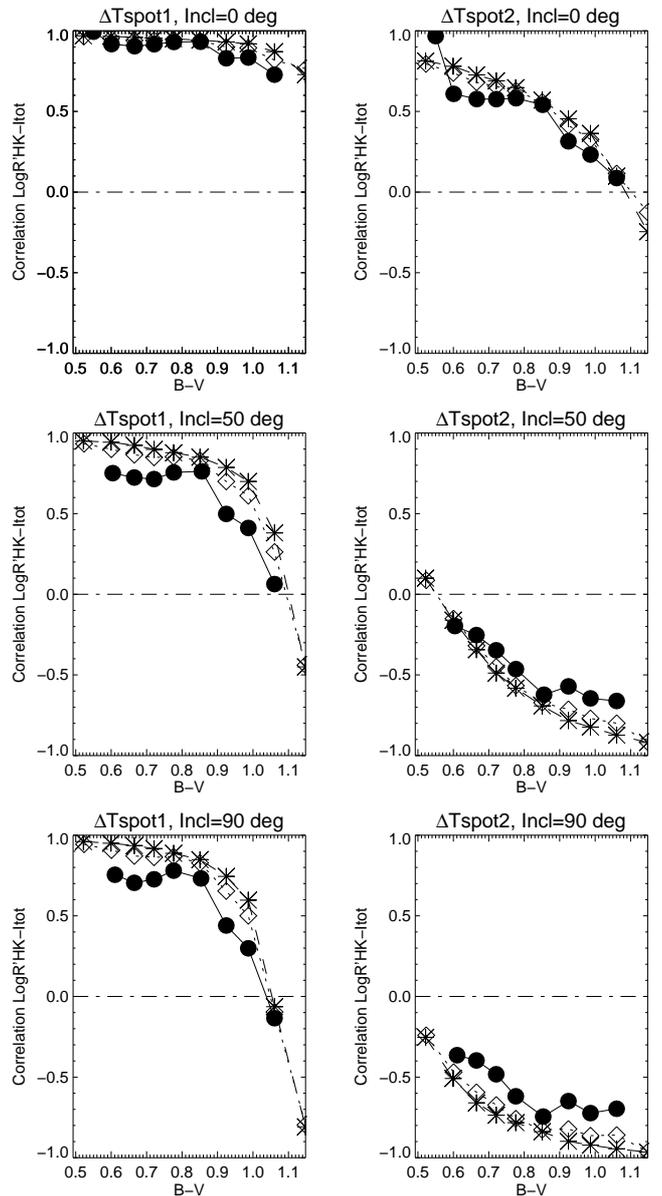}
\caption{
Correlation between LogR'$_{\rm{HK}}$ and I$_{\rm{tot}}$ binned in B-V for $\Delta$Tspot$_1$ (left) and $\Delta$Tspot$_2$ (right), for several activity levels: Average LogR'$_{\rm{HK}}$ below -5.0 (circles), between -4.9 and -4.8 (stars) and above -4.7 (diamonds). 
{\it Upper panels:} for an inclination of 0$^{\circ}$ (pole-on) 
{\it Middle panels:} for an inclination of 50$^{\circ}$.
{\it Lower panels:} for an inclination of 90$^{\circ}$ (edge-on).
}
\label{correl}
\end{figure}

\begin{figure}
\includegraphics*{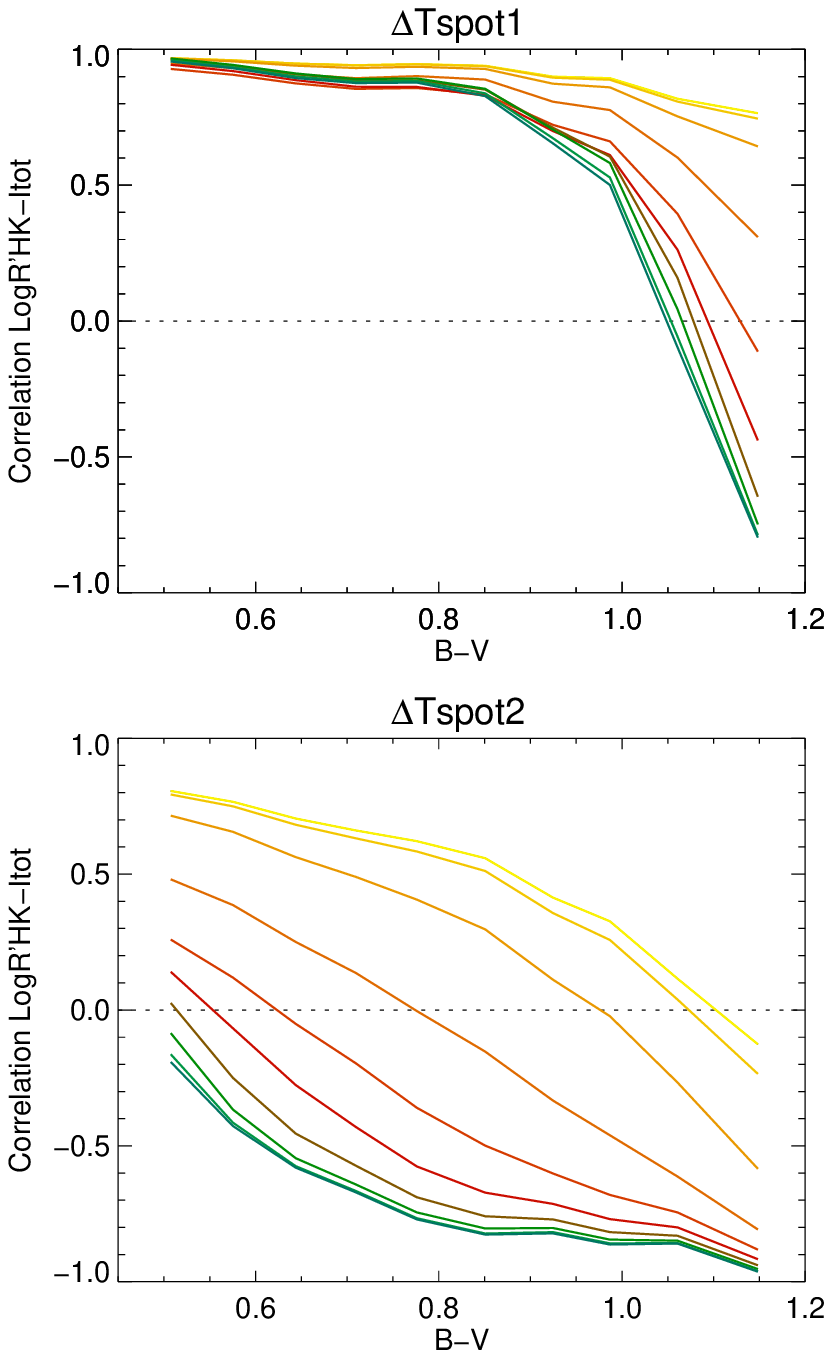}
\caption{
Correlation between LogR'$_{\rm{HK}}$ and I$_{\rm{tot}}$ binned in B-V for $\Delta$Tspot$_1$ (upper panel) and $\Delta$Tspot$_2$ (lower panel) and all stars in grid, for various inclinations. The color-coding corresponds to the inclination, from pole-on (i=0$^{\circ}$, yellow) to edge-on (i=90$^{\circ}$, blue), with light and dark orange corresponding to 20$^{\circ}$ and 30$^{\circ}$, light and dark red to 40$^{\circ}$ and 50$^{\circ}$, brown to  60$^{\circ}$, and light and dark green  to 70$^{\circ}$ and 80$^{\circ}$. 
 }
\label{correl_incl}
\end{figure}

It is very interesting to study the relative contributions of plages and spots on long-term brightness variations. We first present the context for the Sun, and describe how this property has been estimated for other stars in previous works. We apply these techniques to our simulations and check whether they truly correspond to spot-dominated or plage-dominated regimes.

\subsection{Context}

In the solar case, the long-term brightness variations are dominated by plages, whereas spots dominate on rotational timescales \cite[e.g.,][]{shapiro16}. 
Their respective contributions to long-term photometric variations have been measured for a large sample of stars using the correlation between photometric variations and chromospheric emission \cite[][]{radick98,lockwood07,radick18}, based on long-term monitoring in photometry and spectroscopy the Lowell Observatory. The chromospheric emission is related to the plage filling factor.
%, i.e. does not present the degeneracy between plages and spots as the total brightness does. 
They found that for very young and active stars, the correlation is always negative, showing a dominant contribution of spots. For old stars, it is mostly positive (e.g.,  the Sun) and shows a dominant contribution of plages.\ However, there is also a significant number of stars with negative correlations. These older stars are of particular interest here since they correspond to our grid of parameters. 

The Vaughan-Preston gap \cite{vaughan80} separates the two regimes. This has been confirmed by \cite{reinhold19}
on a smaller sample of stars.\ However,  a different approach was used to distinguish between
the two regimes. A phase shift between the two time 
series was computed instead of indicating the correlation between the time series. We note that  the eleven stars in \cite{reinhold19} that correspond 
to our grid parameters (i.e., less active than the Vaughan-Preston gap) show a plage-dominated regime in this paper, while those above the Vaughan-Preston gap are all spot-dominated. Our plage-dominated regime is in agreement with \cite{lockwood07}.
These particular stars are all plage-dominated in \cite{lockwood07}.\ There are only two exceptions of   spot-dominated regimes in this earlier work: one of which has a low degree of significance and positive correlation with the data in Reinhold et al. 2019. The other exception is that of a phase shift at an intermediate value between the two regimes. However, we note 
that while both \cite{lockwood07} and \cite{radick98} observed  a large number  of spot-dominated stars for less active stars than the Vaughan-Preston gap (27-42\%), \cite{reinhold19} observed none.\ It would be interesting to investigate whether this is due to the small subset of stars and their particular selection, or if the different approaches provides different results when the regime is uncertain (see Sect.~6.2.4). 
\cite{radick98}, \cite{lockwood07}, and \cite{radick18} also computed the slope of the brightness variation  versus the chromospheric emission variation (from the S-index), which should provide similar information.

In this section we characterize the correlation between chromospheric emission and brightness time series, the slope, and other indicators from our simulated time series. The objective is three-fold: firstly, to check whether the correlations and slopes derived from our simulations are compatible with observations; secondly, to study the parameter impact on the correlation to better understand what the spot- or plage-dominated regime   means; and thirdly, to study the relation between these observables and the actual contributions of spots and plages to the brightness variation, which are not observables.

\subsection{LogR'$_{\rm{HK}}$-I$_{\rm{tot}}$ correlation and slope}

In this section, we first study the behavior of the correlation and its relationship with R$_{\rm{per}}$. The effect of parameters is studied firstly with $\Delta$Tspot and secondly with the other parameters in our simulation. Finally, we consider  the slope, which is complementary to the correlation, and the effect of seasonnal averages on the results.

\subsubsection{Relation between the correlation and  $\Delta$Tspot}

\begin{figure}
\includegraphics*{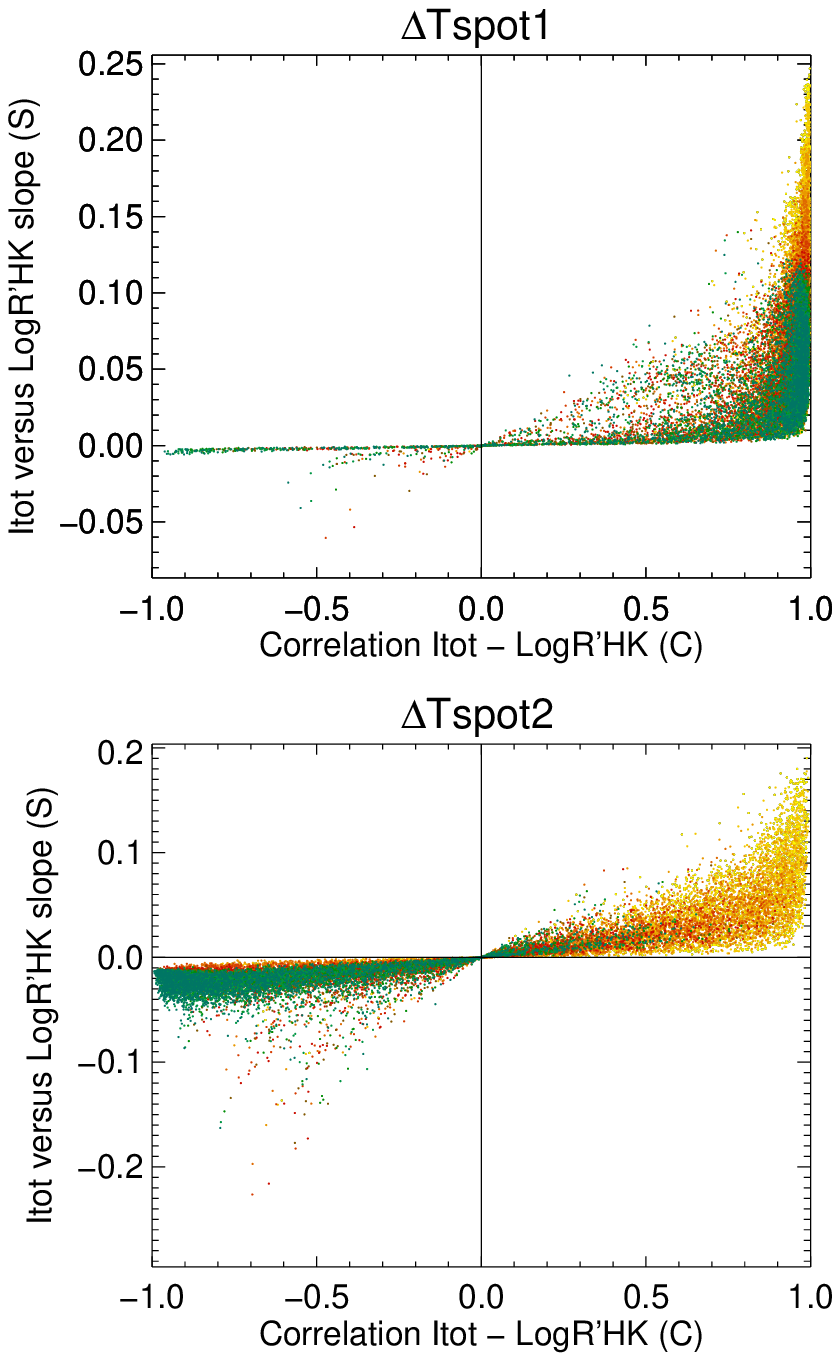}
\caption{
$S$  vs. $C$ for $\Delta$Tspot$_1$ (upper panel) and $\Delta$Tspot$_2$ (lower panel). The color-coding corresponds to the inclination, from pole-on (i=0$^{\circ}$, yellow) to edge-on (i=90$^{\circ}$, blue), with light and dark orange corresponding to 20$^{\circ}$ and 30$^{\circ}$, light and dark red to 40$^{\circ}$ and 50$^{\circ}$, brown to  60$^{\circ}$, and light and dark green  to 70$^{\circ}$ and 80$^{\circ}$. 
Only one simulation out of five is plotted for clarity.
}
\label{slope}
\end{figure}

Our first approach was to compute the correlation $C$ between the long-term variability of LogR'$_{\rm{HK}}$ and I$_{\rm{tot}}$ , either for $\Delta$Tspot$_1$ or $\Delta$Tspot$_2$. For that purpose, the series are smoothed with a binning of one year, and the Pearson correlation $C$ is computed. 
Figure~\ref{ica_typique} shows a typical example of the short- and long-term variability. $C$ is the correlation between the two red curves (from the upper and lower panels, respectively), in this example $C$ is positive and very close to one. 
Figure~\ref{dist_correl} compares the distribution of average R$_{\rm{per}}$ values for simulations in which $C>$0.6 (i.e., strongly correlated and plage dominated) and for simulations in which $C<$-0.6 (i.e., strongly anti-correlated and spot dominated). The plots are not very sensitive to the threshold choice. We find that $\Delta$Tspot$_1$ corresponds to the solar contrast and we observed only correlated time series and no anti-correlated ones, as is observed for the Sun. On the other hand, for $\Delta$Tspot$_2$ with larger contrast, both configurations are present. It is therefore in good agreement with observations \cite[][]{radick98,lockwood07,radick18}, as they observed, both configurations are among old main-sequence solar type stars.

Furthermore, the distributions in Fig.~\ref{dist_correl} show that anti-correlations, when present, are associated with larger R$_{\rm{per}}$ with a threshold around 1000-2000 ppm. 
This is compatible with the results of \cite{montet17} who also found that the two configurations were present from the analysis of the Kepler light curves of solar type stars. 
We selected our simulations corresponding to their the B-V range.  The distributions are on the right-hand side of Fig.~\ref{dist_correl}, which show that the two configurations have  a similar separation in R$_{\rm{per}}$ just as  they obtained: We selected stars within our P$_{\rm{rot}}$ range from their sample.\ We found that 36 stars are plage-dominated, with an average R$_{\rm{per}}$ of 2503 ppm, and 163 stars are spot-dominated, with an average R$_{\rm{per}}$ of 5900 ppm. If we assume that stars may have spot temperatures within the range we have considered, then all these configurations should indeed be possible. 
Overall, for $\Delta$Tspot$_1$, 94\% of the stars from the simulation have $|C|$ larger than 0.5 while 82\% have $|C|$  larger than 0.8. The percentages are only 60\% and 28\% for $\Delta$Tspot$_2$.

% medianes resp 1721 et 5416

\subsubsection{Impact of the parameters on the correlation}

We now investigate, in more detail, how the parameters impact the correlation in order to understand what could be at the origin of this mixed regime. 
We average $C$ in bins of B-V and plot it for various selections of parameters. Figure~\ref{correl} shows the inclination and activity level effects for both values of $\Delta$Tspot. For $\Delta$Tspot$_1$, inclination plays a minimal role, except for K stars for which $C$ decreases when inclination increases. The average activity level seems to have a complex impact with a non-monotonous variation of $C$ with LogR'$_{\rm{HK}}$. There is a strong impact on the spectral type, especially when going toward edge-on configurations. However, for $\Delta$Tspot$_2$, there is a reversal in $C$ with inclination. The anti-correlated time series correspond to close to edge-on configuration, while the same stars viewed pole-on exhibit a positive correlation. Figure~\ref{correl_incl} summarizes the dependence on inclination with a binning in B-V. For $\Delta$Tspot$_1$, the reversal occurs only for K stars, otherwise there is no strong inclination effect. However, for $\Delta$Tspot$_2$, the reversal occurs at a higher inclination when the stellar temperature increases (i.e., there is a larger proportion of spot-dominated light curves). On average, the anti-correlation is present for all inclinations for the lower mass stars in our grid of parameters. 
 The R$_{\rm{per}}$ value is larger for larger inclinations and is thus stronger when it is spot-dominated. This effect was seen by \cite{shapiro14} for an extrapolation of solar time series to more active stars. 

Finally, we looked at the effect of other parameters. The value $C$ is closer to one or negative one when the cycle amplitude is large, while P$_{\rm{cyc}}$ does not impact $C$. This is not illustrated here. The effect of the maximum latitude in the butterfly diagram ($\theta_{\rm{max}}$) is small in most cases, except for $\Delta$Tspot$_2$ and low inclinations. In that case, when $\theta_{\rm{max}}$ increases from solar values, the correlation tends toward zero on average. The $\theta_{\rm{max}}$ value also has a small effect on $\Delta$Tspot$_1$ and low inclinations, but to a lesser extent. 

%Fig.~\ref{correl2} is similar to Fig.~\ref{correl}, 

\subsubsection{Slope properties}

The slope $S$ of I$_{\rm{tot}}$  versus the S-index was also  computed from observations \cite[][]{radick98,lockwood07,radick18}. We show the relationship between $S$ and $C$ from our simulations in Fig.~\ref{slope}. The slopes have values that correspond well to the slopes obtained from observations, typically between -0.25 and 0.25.\ Most observations are between -0.05 and 0.2. A few stars with larger slopes were observed, but there are very few and it is not easy to estimate whether these stars have significantly different behavior or if it is within the uncertainties. A larger dispersion in observations could also be due to metallicity effects \cite[][]{karoff18}, which are not taken into account here. The values $C$ and $S$ have the same sign, so they give the same information on the correlation regime.

\begin{figure}
\includegraphics*{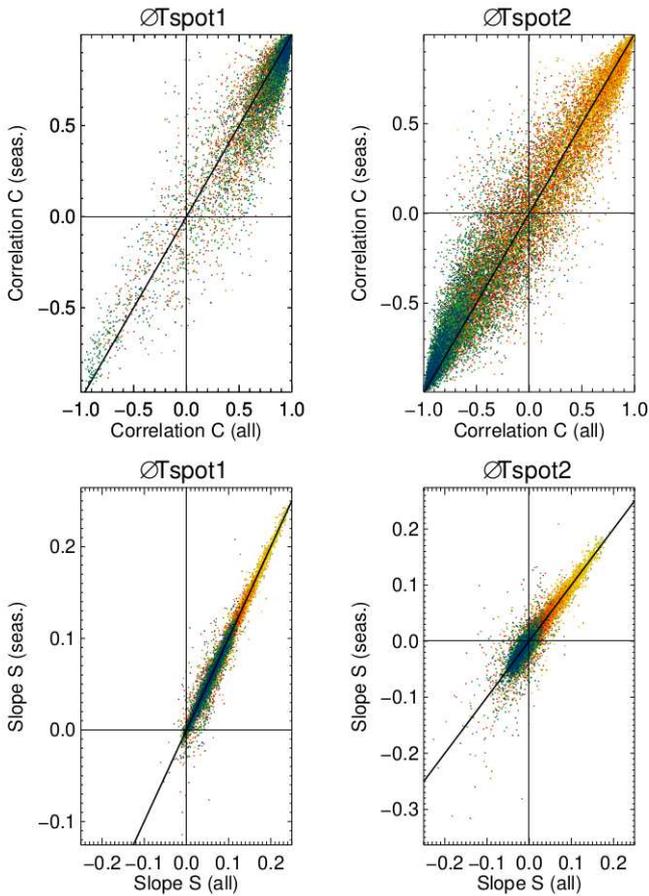}
\caption{
Correlation $C_{50}$ (upper panels) and $S$ (lower panels)  vs. $C$ for $\Delta$Tspot$_1$ (left) and $\Delta$Tspot (right). 
The color-coding corresponds to the inclination, from pole-on (i=0$^{\circ}$, yellow) to edge-on (i=90$^{\circ}$, blue), with light and dark orange corresponding to 20$^{\circ}$ and 30$^{\circ}$, light and dark red to 40$^{\circ}$ and 50$^{\circ}$, brown to  60$^{\circ}$, and light and dark green  to 70$^{\circ}$ and 80$^{\circ}$.  
Only one simulation out of five is plotted for clarity.
}
\label{season}
\end{figure}

\subsubsection{Impact of the sampling on the correlation and slope}

In previous sections, the correlations and slopes were computed from smoothed time series with a very good temporal sampling, which is an ideal case. We now test the impact of the sampling and yearly averaging on the correlation to mimic the observing conditions of \cite{radick98}. We assume a gap in observations of four months per year. For each year, we randomly selected either 30, 50, or 70 ($N$) points on the remaining dates, and then computed the average of the LogR'$_{\rm{HK}}$ for each year, the total brightness, and S-index. We then recalculated these correlations (hereafter $C_N$) and the slope ($S_N$) from these seasonal time series. The objective was to compare them with a correlation and slope obtained with a very good temporal sampling ($C$ as studied above). Figure~\ref{season} shows the results for N=50 (N=30 and 70 are similar, with a slightly larger dispersion for 30 and a slightly lower dispersion for 70).
There is a good correlation between $C$ and $C_N$ (Pearson correlation of 0.95), but also a large dispersion. When selecting points with $C$ above 0.8, the dispersion decreases from 0.056 to 0.039 (for $N$ from 30 to 70), indicating the improvement brought by the increasing $N$. The dispersion is slightly larger for $C$ below -0.8 (from 0.10 to 0.08). However, in the range of $C$ between -0.3 and 0.3, for example, the dispersion is on the order of 0.25. Therefore, these correlations are poorly determined and their sign is meaningless: Only strong correlations can be truly representative of the plage- or spot- dominated regime. The same is true for the slopes (lower panel of Fig.~\ref{season}), which are not reliable for absolute values below 0.02-0.03, although the dispersion is lower than for $C$. 
We conclude that the correlations or slopes depend on the sampling. It is interesting to note that only
the strong correlations or slopes are significant with a yearly averaging the temporal series and limited number of points.

%%%%We are therefore able to reproduce some properties similar to observations, by only changing the spot contrast within a reasonable range (the spot/plage size ratio distribution used to generate them follows the same distribution for all series). However many parameters impact the correlations, and in particular in some configurations it may not be intrinsic to the star, and may depend on the point of view. Care must also be taken about low correlations and slopes, whose signs are not reliable then. 

\subsection{Relation between $C,$  actual plage, and spot intensity contribution}

\begin{figure}
\includegraphics*{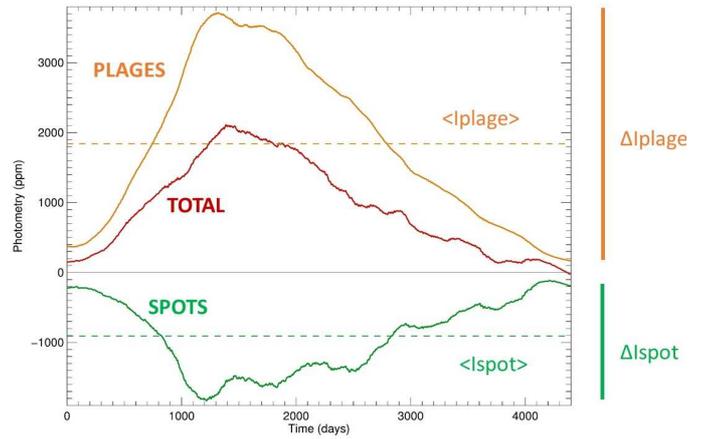}
\caption{
Typical long-term brightness variation due to plages (orange line). The green curve is for spots, and the red curve represents the sum of the two. The two horizontal lines correspond to the average (used to compute $R_I$), while the vertical bars on the right side are the long-term amplitudes (used to compute $R_A$). 
}
\label{schema}
\end{figure}

\begin{figure}
\includegraphics*{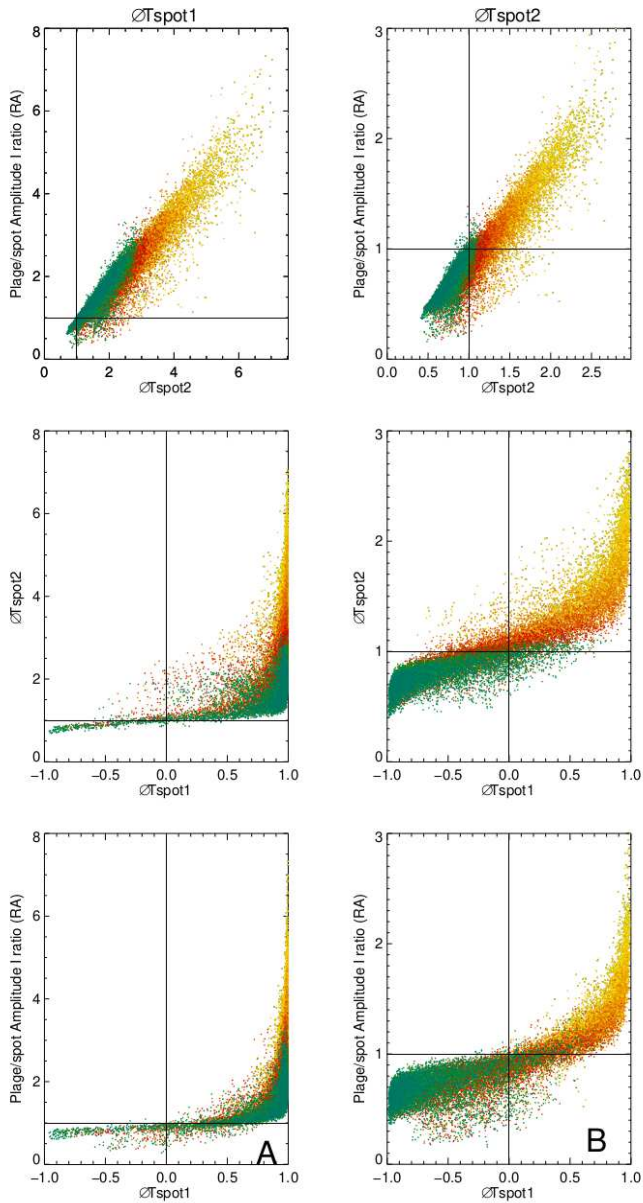}
\caption{
$R_A$  vs. $R_I$ (upper panels), $R_I$  vs. $C$ (middle panels), and $R_A$  vs. $C$ (lower panels), for $\Delta$Tspot$_1$ (left panel) and $\Delta$Tspot$_2$ (right panel). 
The color-coding corresponds to the inclination, from pole-on (i=0$^{\circ}$, yellow) to edge-on (i=90$^{\circ}$, blue), with light and dark orange corresponding to 20$^{\circ}$ and 30$^{\circ}$, light and dark red to 40$^{\circ}$ and 50$^{\circ}$, brown to  60$^{\circ}$, and light and dark green  to 70$^{\circ}$ and 80$^{\circ}$. 
Only one simulation out of five is plotted for clarity.
}
\label{crit}
\end{figure}

%Letters A to F indicates quadrants (with respect to the black vertical and horizontal lines) corresponding to simulations where the criteria provides opposite informations (see text). 

The correlation $C$ and the slope $S$ can be computed from observations, and are used as proxies to determine if plages or spots dominate the brightness variations. These individual contributions can not be extracted from observations because the brightness is the residual between the two contributions. However, we can use our large set of simulated configurations to check how these proxies relate to actual plage and spot brightness contributions. For that purpose, we define two  criteria illustrated in Fig.~\ref{schema}. 
Firstly, R$_A$ is the ratio between $\Delta$I$_{\rm{plage}}$  and $\Delta$I$_{\rm{spot}}$, which are the amplitudes of I$_{\rm{plage}}$ and I$_{\rm{spot}}$ and relates to the contribution of spots and plages to the variability.\ It is important to note that after the amplitudes are smoothed, the long-term amplitude can be calculated, which is defined as the maximum
minus the minimum for each time series),   
Secondly, $R_I$ is the ratio between the temporally averaged I$_{\rm{plage}}$ (i.e., $<$I$_{\rm{plage}}$$>$, representative of the total flux corresponding to plages) and the temporally averaged I$_{\rm{spot}}$ (i.e., $<$I$_{\rm{spot}}$$>$).\ This relates to the total amount of flux in spots and plages, respectively.  
 It is not possible to  directly observe $R_A$ and $R_I$. 
The red curve in Fig.~\ref{schema} is the one which is used together with LogR'$_{\rm{HK}}$  in the previous  section to estimate the correlation $C$. 

Figure~\ref{crit} shows the relationship between $C$, $R_A$, and $R_I$. 
Both $C$ (correlation between the photometric and chromospheric time series) and $R_A$ (signed ratio between the $\Delta$I$_{\rm{plage}}$  and $\Delta$I$_{\rm{spot}}$) allow us to identify whether plages dominate the brightness variability or not. We therefore expect $C$ to be positive when $R_A$ is larger than one, and negative when  $R_A$ is lower than one.
The same is true for $R_I$ if it is larger than one when plages provide a larger flux on average than spots. 
For many simulations, this is indeed the case as most points are either in the upper right corner or the lower left corner of each plot. For example, for the $C_1$--$R_{A1}$ relation, 97\% of the points are in these quadrants for $\Delta$Tspot$_1$ and 90\% for $\Delta$Tspot$_2$. 

However, two other major features can be seen. Firstly, there are a significant number of stars in quadrants where we would not expect them if $C$ is indeed representative of the respective plage and spot contributions as described above. All scatter plots show this behavior. We concentrate on the $C$-$R_A$ relationship here, as $R_A$ is related to the variability. The upper left quadrant is usually not populated, but about 4\% and 10\% are in the lower right quadrants for $\Delta$Tspot$_1$ and $\Delta$Tspot$_2  $respectively.\ This is illustrated in quadrants A and B  in Fig.~\ref{crit} in which $C$ is positive, which would correspond to plage-dominated.\ Additionally, $R_A$ is lower than one, which would correspond to spot-dominated. They are spread in spectral type, inclination, and activity level. The most extreme cases tend to correspond to low activity stars close to edge-on. 
These departures can occasionally be due to short-term effects caused by very large spot contribution, which is a rare. In any case, this dispersion adds to the uncertainties already discussed in the previous section and leads to the conclusion that correlation below 0.5-0.6 may not be representative of the actual sign.

Secondly, there is a strong inclination effect, for example large values of $R_I$ are seen only at low inclinations. The trend is similar for  $R_A$. 
Stars seen edge-on tend to have $R_A$ or $R_I$ closer to one than stars seen pole-on. Therefore, a star like the Sun will have specific behavior in terms of photometric variability \cite[as proposed by][]{schatten93} as studied by \cite{knaack01} and \cite{shapiro14}, because the Sun tends to have a low photometric variability compared to its chromospheric amplitude and average activity level \cite[e.g.,][]{radick98}. 
Consequently, it is beyond the scope of this paper to study the solar case in detail.\  Unlike \cite{knaack01} or \cite{shapiro14}, who studied the wavelength effect on the photometry, the plage contrast used in our paper does not correspond to these solar observations \cite[see also the discussions in][]{shapiro15,shapiro16,radick18}. 
However, it is interesting to point out that the relative difference we obtained between the long-term photometric amplitude for a star seen edge-on and the same star seen pole-on can be quite large.\ This strongly depends on $\Delta$Tspot and can also vary significantly from one simulation to the other, as studied in Sect.~3.2.

\subsection{Comparison with radial velocities}

\begin{figure}
\includegraphics*{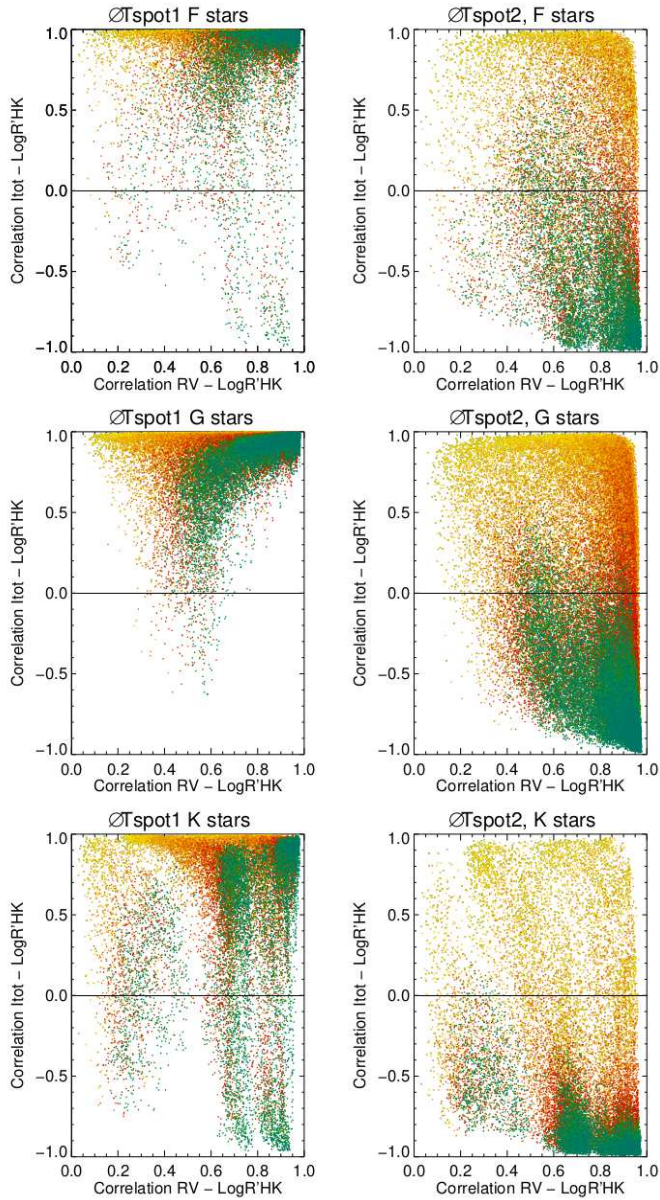}
\caption{
Global correlation between LogR'$_{\rm{HK}}$ and I$_{\rm{tot}}$  vs. correlation between RV and LogR'$_{\rm{HK}}$ for F stars (upper panels), G stars (middle panels), and K stars (lower panels), for $\Delta$Tspot$_1$ (left) and $\Delta$Tspot$_2$ (right). The RV time series include activity, the signal due to oscillations, granulation, and supergranulation averaged over 1 hour and an instrumental white noise of 0.6 m/s. The color-coding corresponds to the inclination, from pole-on (i=0$^{\circ}$, yellow) to edge-on (i=90$^{\circ}$, blue), with light and dark orange corresponding to 20$^{\circ}$ and 30$^{\circ}$, light and dark red to 40$^{\circ}$ and 50$^{\circ}$, brown to  60$^{\circ}$, and light and dark green  to 70$^{\circ}$ and 80$^{\circ}$. 
Only one simulation out of five is plotted for clarity.
}
\label{rv}
\end{figure}

As seen in both observations and simulations, it is possible to find stars which are either spot-dominated or plage-dominated in brightness variation across our grid. Here, we compare the correlations for brightness ($C$) with the correlations between RV time series and LogR'$_{\rm{HK}}$. The time series RV are dominated in large part by the inhibition of convective blueshift in plages, and therefore by plages. Figure~\ref{rv} shows the brightness correlations  versus the RV correlations for the two $\Delta$Tspot and different spectral types. 

The two observables, photometry and RV, clearly behave differently. There is a large proportion of simulations with strong correlations, but while the RV correlation is always positive, the brightness correlation can have either sign. It is therefore possible to have a star that is plage-dominated in RV but spot-dominated in its brightness variations. This is mainly observed for $\Delta$Tspot$_2$ (see Sect.\ 5), but it is also possible for $\Delta$Tspot$_1$ for K stars. This means that if the contribution
of the convective blueshift inhibition is well removed for these configurations, then the residuals will probably be dominated by spots. 
It is also  possible to have a star with a low RV-LogR'$_{\rm{HK}}$ correlation but a strong I$_{\rm{tot}}$-LogR'$_{\rm{HK}}$ correlation, especially for edge-on configurations. 
Inclination plays an important role. This is mostly seen for the  I$_{\rm{tot}}$-LogR'$_{\rm{HK}}$ correlations. However, it is present to a lesser extent for the RV-LogR'$_{\rm{HK}}$ correlation and seen mostly for stars with a good I$_{\rm{tot}}$-LogR'$_{\rm{HK}}$ correlation in the $\Delta$Tspot$_1$ case. 
Edge-on stars are those with the higher RV-LogR'$_{\rm{HK}}$ correlation in all cases, but also those with the largest dispersion in the I$_{\rm{tot}}$-LogR'$_{\rm{HK}}$ correlation distribution. 

The notion of spot-dominated and plage-dominated regime is, therefore, strongly dependent on the observables.

%-----------------------------------
%-----------------------------------
\section{Conclusion}
%-----------------------------------
%-----------------------------------

The analysis of a large number of complex and realistic  simulations of brightness and LogR'$_{\rm{HK}}$ time series allowed us to make predictions on the variability for a large domain of stellar parameters and to provide some clues and limitations about the interpretation of observed stellar light  curves. 
For old main-sequence stars, the short-term variability R$_{\rm{per}}$ , which is computed over 90 days as in Kepler data and defined as the amplitude within the 5\%-95\% percentiles, is found to increase with decreasing T$_{\rm{eff}}$.\  This was observed by  Kepler; however, with a lower level and a lower slope. 
Even if the trend is similar, the precise comparison with observations is difficult because of the biases in observations,  the distribution of the parameters in our simulations is not known, and the necessity to compare similar samples. For example, the question is begged of whether there are more stars with a low spot contrast, larger spot contrast, and is this changes with T$_{\rm{eff}}$ . It is also difficult to measure the rotation period of stars with low inclinations, and, therefore, low short-term variability. A low R$_{\rm{per}}$ and unreliable P$_{\rm{rot}}$  does not mean that the star is intrinsically quiet. A short lifetime compared to a long rotation period also prevents a good measurement due to the loss of coherence.  

The value R$_{\rm{per}}$ is correlated with LogR'$_{\rm{HK}}$, but with a very large dispersion preventing the formation of a very precise relationship between these two observables. In our simulations, this is mostly due to inclination and spot contrast. It is important to be very careful
when comparing laws from various sources unless these effects can be mitigated by a better knowledge of the star. It may also be difficult to distinguish between low inclination high activity stars and high inclination low activity stars. The same is true for the relationship between the short-term variability R$_{\rm{per}}$ and long-term variability, which have a strong dependence on inclination and spot contrast.\  This shows a complex pattern. 
We also found that the long-term brightness variations between edge-on and pole-on orientations presented a large diversity of situations. They often have the same sign with different signs of variations and ratios between amplitudes, which can be higher or lower than one.\ However, they can also have a different sign.\ This shows more complex behavior than the decrease of the long-term variability when inclination increases for a solar-like patter, as found by \cite{shapiro14}. 
Although there is a clear relationship between the short-term photometric variability R$_{\rm{per}}$ and the RV jitter, this relationship strongly depends on the type of stars and its properties: spectral type, P$_{\rm{rot}}$, spot properties, and inclination. The dispersion is such that it would be unwise to use one simple law to then make all predictions. However, the range of possible values can be narrowed down if the spectral type, P$_{\rm{rot}}$  , and inclination are well constrained. Nevertheless, the uncertainty due to the spot contrast will probably remain. The reconstruction of radial velocities from photometry as done in \cite{aigrain12} is therefore likely to be complex when the activity pattern is as complex as studied here. 

The number of structures controls the variability in our simulations, illustrating the strong degeneracy between the size of structure, their contrast, and their number. This prevents any direct interpretation of the observed variability in terms of spot size for example.

We performed an extensive analysis of the correlation between the brightness variations and LogR'$_{\rm{HK}}$, which were both used as a proxy to determine whether a star brightness variation was dominated by plages or spots. We discovered that we were able to reproduce the observations by using a sufficiently large spot contrast, which is compatible with what
we currently know of stellar spot temperatures. It is important to mention that the observations were mixed regime between spot-dominated and plage-dominated for old stars, spot dominated stars with a larger R$_{\rm{per}}$, and slope of brightness  versus LogR'$_{\rm{HK}}$. Furthermore, the existence of the two regimes seems to be due to inclination, a parameter that is not intrinsic to the star. This is also related to the strong relation between apparent variability and inclination. Different dynamo models are, for example, not necessary to explain these observations. \cite{shapiro14} also found that a reversal in regime could be due to inclination when simply extending the solar pattern to much more active stars. In the present paper, we study this behavior in detail as a function of spectral type since the effect is more pronounced for K stars as compared to F stars.\ This dependence may explain the similar trend observed for the boundary between plage-dominated and spot-dominated stars by \cite{radick98}, \cite{lockwood07}, \cite{radick18}, and \cite{reinhold19}, as spot-dominated regime are also reached at lower activity levels for lower mass stars. \cite{witzke18} also found a change in metallicity, which was not taken into account here.  This then results in a change in plage
contrast, which could also lead to different behaviors (spot- or plage-dominated
depending on the star metallicity).\ The different behaviors play a role, similar to our dispersion in spot contrast,  and both are probably at play. However, we find that inclination must play a strong role.   

Furthermore, we checked whether this correlation, which is used as a proxy to determine if the long-term brightness variations are spot- or plage- dominated, actually corresponds to this assumption. We  analyzed the amplitude of brightness variations due to plages and spots separately, in addition to their average level. Up to 10\% of the simulations (for $\Delta$Tspot$_2$) correspond to cases where the correlation is positive but the actual variability is dominated by spots. At these low levels of correlation, the correlation between the time series is sensitive to the occasional presence of large spots, even with the one year smoothing.  
In addition, correlations and slopes computed from seasonal averages based on a limited number of points exhibit a large dispersion.\ The result is that absolute correlations lower than 0.5-0.6, or slope within 0.02-0.03, do not have a reliable sign. 
Overall, the correlation between LogR'$_{\rm{HK}}$ and the total long-term brightness variation is representative of the true plage- or spot-dominated regime only when it is large, in absolute value.\ This is because weak correlations may correspond to different situations and are also strongly impacted by a poor sampling. 
 
%%%%SUMMARY : $C$ behaves has observed (miwed regime, dependence on R$_{\rm{per}}$). However it is not entirely intrinsic ($C$ may change sign for a given star, with inclination). And in low variability stars, the correspondence with actual relative contribution of plages and spots to brightness variability not always valid. 

Finally, we insist on the major role  inclination plays on all the properties we have studied in this paper, such as R$_{\rm{per}}$, relation with LogR'$_{\rm{HK}}$, and plage- or spot-dominated regime. Inclination usually impacts the characteristics of these effects in conjunction with other geometrical effects, such as the coverage in latitude. Inclination is not intrinsic to the star, and, therefore, is not related to the dynamo at play for example.\ However, it is often overlooked when interpreting brightness variations. 
The notion of spot-dominated and plage-dominated regime is therefore strongly dependent on the observables and should not be used in a general manner. 

%%%%\textbf{ lien photmo-RV ; si selection etoile obs photom sur base petit Rper / pour follow-up + facile, risque de tomber sur inclinaison elevee = pas favorable du tout du point de vue exopla)

%%%%%%%%%%%%%%%%%%%%%%%%%%%%%%%%%%%%%%%%%%%%%%%%%%%%%
%%%%%%%%%%%%%%%%%%%%%%%%%%%%%%%%%%%%%%%%%%%%%%%%%%%%%
\begin{acknowledgements}

This work has been funded by the ANR GIPSE ANR-14-CE33-0018.
This work was supported by the Programme National de Physique Stellaire (PNPS) of the CNRS/INSU, co-funded by the CEA, the CNES, and the Programme National de Plan\'etologie (PNP) of the CNRS/INSU, co-funded by the CNES.
We are very grateful to Charlotte Norris, who has provided us with the plage contrasts used in this work prior to the publication of her thesis.

\end{acknowledgements}

%%%%%%%%%%%%%%%%%%%%%%%%%%%%%%%%%%%%%%%%%%%%%%%%%%%%%
%%%%%%%%%%%%%%%%%%%%%%%%%%%%%%%%%%%%%%%%%%%%%%%%%%%%%
\bibliographystyle{aa}
\bibliography{biblio}

\begin{thebibliography}{53}
\expandafter\ifx\csname natexlab\endcsname\relax\def\natexlab#1{#1}\fi

\bibitem[{{Aigrain} {et~al.}(2012){Aigrain}, {Pont}, \& {Zucker}}]{aigrain12}
{Aigrain}, S., {Pont}, F., \& {Zucker}, S. 2012, \mnras, 419, 3147

\bibitem[{{Arkhypov} {et~al.}(2015){Arkhypov}, {Khodachenko}, {Lammer},
  {G{\"u}del}, {L{\"u}ftinger}, \& {Johnstone}}]{arkhypov15}
{Arkhypov}, O.~V., {Khodachenko}, M.~L., {Lammer}, H., {et~al.} 2015, \apj,
  807, 109

\bibitem[{{Baliunas} {et~al.}(1995){Baliunas}, {Donahue}, {Soon}, {Horne},
  {Frazer}, {Woodard-Eklund}, {Bradford}, {Rao}, {Wilson}, {Zhang}, {Bennett},
  {Briggs}, {Carroll}, {Duncan}, {Figueroa}, {Lanning}, {Misch}, {Mueller},
  {Noyes}, {Poppe}, {Porter}, {Robinson}, {Russell}, {Shelton}, {Soyumer},
  {Vaughan}, \& {Whitney}}]{baliunas95}
{Baliunas}, S.~L., {Donahue}, R.~A., {Soon}, W.~H., {et~al.} 1995, \apj, 438,
  269

\bibitem[{{Basri} {et~al.}(2011){Basri}, {Walkowicz}, {Batalha}, {Gilliland},
  {Jenkins}, {Borucki}, {Koch}, {Caldwell}, {Dupree}, {Latham}, {Marcy},
  {Meibom}, \& {Brown}}]{basri11}
{Basri}, G., {Walkowicz}, L.~M., {Batalha}, N., {et~al.} 2011, \aj, 141, 20

\bibitem[{{Basri} {et~al.}(2010){Basri}, {Walkowicz}, {Batalha}, {Gilliland},
  {Jenkins}, {Borucki}, {Koch}, {Caldwell}, {Dupree}, {Latham}, {Meibom},
  {Howell}, \& {Brown}}]{basri10}
{Basri}, G., {Walkowicz}, L.~M., {Batalha}, N., {et~al.} 2010, \apjl, 713, L155

\bibitem[{{Berdyugina}(2005)}]{berd05}
{Berdyugina}, S.~V. 2005, Living Reviews in Solar Physics, 2, 8

\bibitem[{{Bonomo} \& {Lanza}(2008)}]{bonomo08}
{Bonomo}, A.~S. \& {Lanza}, A.~F. 2008, \aap, 482, 341

\bibitem[{{Borgniet} {et~al.}(2015){Borgniet}, {Meunier}, \&
  {Lagrange}}]{borgniet15}
{Borgniet}, S., {Meunier}, N., \& {Lagrange}, A.-M. 2015, \aap, 581, A133

\bibitem[{{Davenport}(2017)}]{davenport17}
{Davenport}, J.~R.~A. 2017, \apj, 835, 16

\bibitem[{{Ferreira Lopes} {et~al.}(2015){Ferreira Lopes}, {Le{\~a}o}, {de
  Freitas}, {Canto Martins}, {Catelan}, \& {De Medeiros}}]{ferreiralopes15}
{Ferreira Lopes}, C.~E., {Le{\~a}o}, I.~C., {de Freitas}, D.~B., {et~al.} 2015,
  \aap, 583, A134

\bibitem[{{Giles} {et~al.}(2017){Giles}, {Collier Cameron}, \&
  {Haywood}}]{giles17}
{Giles}, H.~A.~C., {Collier Cameron}, A., \& {Haywood}, R.~D. 2017, \mnras,
  472, 1618

\bibitem[{{Gray} {et~al.}(2006){Gray}, {Corbally}, {Garrison}, {McFadden},
  {Bubar}, {McGahee}, {O'Donoghue}, \& {Knox}}]{gray06}
{Gray}, R.~O., {Corbally}, C.~J., {Garrison}, R.~F., {et~al.} 2006, \aj, 132,
  161

\bibitem[{{Gray} {et~al.}(2003){Gray}, {Corbally}, {Garrison}, {McFadden}, \&
  {Robinson}}]{gray03}
{Gray}, R.~O., {Corbally}, C.~J., {Garrison}, R.~F., {McFadden}, M.~T., \&
  {Robinson}, P.~E. 2003, \aj, 126, 2048

\bibitem[{{Hall} {et~al.}(2009){Hall}, {Henry}, {Lockwood}, {Skiff}, \&
  {Saar}}]{hall09}
{Hall}, J.~C., {Henry}, G.~W., {Lockwood}, G.~W., {Skiff}, B.~A., \& {Saar},
  S.~H. 2009, \aj, 138, 312

\bibitem[{{He} {et~al.}(2015){He}, {Wang}, \& {Yun}}]{he15}
{He}, H., {Wang}, H., \& {Yun}, D. 2015, \apjs, 221, 18

\bibitem[{{Hempelmann} {et~al.}(2016){Hempelmann}, {Mittag}, {Gonzalez-Perez},
  {Schmitt}, {Schr{\"o}der}, \& {Rauw}}]{hempelmann16}
{Hempelmann}, A., {Mittag}, M., {Gonzalez-Perez}, J.~N., {et~al.} 2016, \aap,
  586, A14

\bibitem[{{Herrero} {et~al.}(2016){Herrero}, {Ribas}, {Jordi}, {Morales},
  {Perger}, \& {Rosich}}]{herrero16}
{Herrero}, E., {Ribas}, I., {Jordi}, C., {et~al.} 2016, \aap, 586, A131

\bibitem[{{Juvan} {et~al.}(2018){Juvan}, {Lendl}, {Cubillos}, {Fossati},
  {Tregloan-Reed}, {Lammer}, {Guenther}, \& {Hanslmeier}}]{juvan18}
{Juvan}, I.~G., {Lendl}, M., {Cubillos}, P.~E., {et~al.} 2018, \aap, 610, A15

\bibitem[{{Karoff} {et~al.}(2018){Karoff}, {Metcalfe}, {Santos}, {Montet},
  {Isaacson}, {Witzke}, {Shapiro}, {Mathur}, {Davies}, {Lund}, {Garcia},
  {Brun}, {Salabert}, {Avelino}, {van Saders}, {Egeland}, {Cunha}, {Campante},
  {Chaplin}, {Krivova}, {Solanki}, {Stritzinger}, \& {Knudsen}}]{karoff18}
{Karoff}, C., {Metcalfe}, T.~S., {Santos}, {\^A}.~R.~G., {et~al.} 2018, \apj,
  852, 46

\bibitem[{{Kipping}(2012)}]{kipping12}
{Kipping}, D.~M. 2012, \mnras, 427, 2487

\bibitem[{{Knaack} {et~al.}(2001){Knaack}, {Fligge}, {Solanki}, \&
  {Unruh}}]{knaack01}
{Knaack}, R., {Fligge}, M., {Solanki}, S.~K., \& {Unruh}, Y.~C. 2001, \aap,
  376, 1080

\bibitem[{{Lanza} {et~al.}(2007){Lanza}, {Bonomo}, \& {Rodon{\`o}}}]{lanza07}
{Lanza}, A.~F., {Bonomo}, A.~S., \& {Rodon{\`o}}, M. 2007, \aap, 464, 741

\bibitem[{{Lanza} {et~al.}(2014){Lanza}, {Das Chagas}, \& {De
  Medeiros}}]{lanza14}
{Lanza}, A.~F., {Das Chagas}, M.~L., \& {De Medeiros}, J.~R. 2014, \aap, 564,
  A50

\bibitem[{{Lanza} {et~al.}(2009){Lanza}, {Pagano}, {Leto}, {Messina},
  {Aigrain}, {Alonso}, {Auvergne}, {Baglin}, {Barge}, {Bonomo}, {Boumier},
  {Collier Cameron}, {Comparato}, {Cutispoto}, {de Medeiros}, {Foing},
  {Kaiser}, {Moutou}, {Parihar}, {Silva-Valio}, \& {Weiss}}]{lanza09}
{Lanza}, A.~F., {Pagano}, I., {Leto}, G., {et~al.} 2009, \aap, 493, 193

\bibitem[{{Lockwood} {et~al.}(2007){Lockwood}, {Skiff}, {Henry}, {Henry},
  {Radick}, {Baliunas}, {Donahue}, \& {Soon}}]{lockwood07}
{Lockwood}, G.~W., {Skiff}, B.~A., {Henry}, G.~W., {et~al.} 2007, \apjs, 171,
  260

\bibitem[{{Mamajek} \& {Hillenbrand}(2008)}]{mamajek08}
{Mamajek}, E.~E. \& {Hillenbrand}, L.~A. 2008, \apj, 687, 1264

\bibitem[{{McQuillan} {et~al.}(2013){McQuillan}, {Aigrain}, \&
  {Mazeh}}]{mcquillan13b}
{McQuillan}, A., {Aigrain}, S., \& {Mazeh}, T. 2013, \mnras, 432, 1203

\bibitem[{{McQuillan} {et~al.}(2014){McQuillan}, {Mazeh}, \&
  {Aigrain}}]{mcquillan14}
{McQuillan}, A., {Mazeh}, T., \& {Aigrain}, S. 2014, \apjs, 211, 24

\bibitem[{{Mehrabi} {et~al.}(2017){Mehrabi}, {He}, \&
  {Khosroshahi}}]{mehrabi17}
{Mehrabi}, A., {He}, H., \& {Khosroshahi}, H. 2017, \apj, 834, 207

\bibitem[{{Meunier} \& {Lagrange}(2019)}]{meunier19b}
{Meunier}, N. \& {Lagrange}, A.~M. 2019, \aap, 628, A125

\bibitem[{{Meunier} {et~al.}(2019){Meunier}, {Lagrange}, {Boulet}, \&
  {Borgniet}}]{meunier19}
{Meunier}, N., {Lagrange}, A.~M., {Boulet}, T., \& {Borgniet}, S. 2019, \aap,
  627, A56

\bibitem[{Meunier {et~al.}(2019)Meunier, Lagrange, \& Cuzacq}]{meunier19c}
Meunier, N., Lagrange, A.-M., \& Cuzacq, S. 2019, submitted to A\&A

\bibitem[{{Mittag} {et~al.}(2017){Mittag}, {Hempelmann}, {Schmitt},
  {Fuhrmeister}, {Gonz{\'a}lez-P{\'e}rez}, \& {Schr{\"o}der}}]{mittag17}
{Mittag}, M., {Hempelmann}, A., {Schmitt}, J.~H.~M.~M., {et~al.} 2017, \aap,
  607, A87

\bibitem[{{Mittag} {et~al.}(2013){Mittag}, {Schmitt}, \&
  {Schr{\"o}der}}]{mittag13}
{Mittag}, M., {Schmitt}, J.~H.~M.~M., \& {Schr{\"o}der}, K.-P. 2013, \aap, 549,
  A117

\bibitem[{{Montet} {et~al.}(2017){Montet}, {Tovar}, \&
  {Foreman-Mackey}}]{montet17}
{Montet}, B.~T., {Tovar}, G., \& {Foreman-Mackey}, D. 2017, \apj, 851, 116

\bibitem[{{Mosser} {et~al.}(2009){Mosser}, {Baudin}, {Lanza}, {Hulot},
  {Catala}, {Baglin}, \& {Auvergne}}]{mosser09}
{Mosser}, B., {Baudin}, F., {Lanza}, A.~F., {et~al.} 2009, ArXiv e-prints

\bibitem[{{Nielsen} {et~al.}(2013){Nielsen}, {Gizon}, {Schunker}, \&
  {Karoff}}]{nielsen13}
{Nielsen}, M.~B., {Gizon}, L., {Schunker}, H., \& {Karoff}, C. 2013, \aap, 557,
  L10

\bibitem[{{Norris}(2018)}]{norris18}
{Norris}, C. 2018, PhD thesis, Imperial College London

\bibitem[{{Noyes} {et~al.}(1984{\natexlab{a}}){Noyes}, {Hartmann}, {Baliunas},
  {Duncan}, \& {Vaughan}}]{noyes84}
{Noyes}, R.~W., {Hartmann}, L.~W., {Baliunas}, S.~L., {Duncan}, D.~K., \&
  {Vaughan}, A.~H. 1984{\natexlab{a}}, \apj, 279, 763

\bibitem[{{Noyes} {et~al.}(1984{\natexlab{b}}){Noyes}, {Weiss}, \&
  {Vaughan}}]{noyes84b}
{Noyes}, R.~W., {Weiss}, N.~O., \& {Vaughan}, A.~H. 1984{\natexlab{b}}, \apj,
  287, 769

\bibitem[{{Radick} {et~al.}(2018){Radick}, {Lockwood}, {Henry}, {Hall}, \&
  {Pevtsov}}]{radick18}
{Radick}, R.~R., {Lockwood}, G.~W., {Henry}, G.~W., {Hall}, J.~C., \&
  {Pevtsov}, A.~A. 2018, \apj, 855, 75

\bibitem[{{Radick} {et~al.}(1998){Radick}, {Lockwood}, {Skiff}, \&
  {Baliunas}}]{radick98}
{Radick}, R.~R., {Lockwood}, G.~W., {Skiff}, B.~A., \& {Baliunas}, S.~L. 1998,
  \apjs, 118, 239

\bibitem[{{Reinhold} {et~al.}(2019){Reinhold}, {Bell}, {Kuszlewicz}, {Hekker},
  \& {Shapiro}}]{reinhold19}
{Reinhold}, T., {Bell}, K.~J., {Kuszlewicz}, J., {Hekker}, S., \& {Shapiro},
  A.~I. 2019, \aap, 621, A21

\bibitem[{{Reinhold} {et~al.}(2017){Reinhold}, {Cameron}, \&
  {Gizon}}]{reinhold17}
{Reinhold}, T., {Cameron}, R.~H., \& {Gizon}, L. 2017, \aap, 603, A52

\bibitem[{{Reinhold} \& {Reiners}(2013)}]{reinhold13}
{Reinhold}, T. \& {Reiners}, A. 2013, \aap, 557, A11

\bibitem[{{Reinhold} {et~al.}(2013){Reinhold}, {Reiners}, \&
  {Basri}}]{reinhold13b}
{Reinhold}, T., {Reiners}, A., \& {Basri}, G. 2013, \aap, 560, A4

\bibitem[{{Saar} \& {Brandenburg}(1999)}]{saar99}
{Saar}, S.~H. \& {Brandenburg}, A. 1999, \apj, 524, 295

\bibitem[{{Schatten}(1993)}]{schatten93}
{Schatten}, K.~H. 1993, \jgr, 98, 18

\bibitem[{{Shapiro} {et~al.}(2014){Shapiro}, {Solanki}, {Krivova}, {Schmutz},
  {Ball}, {Knaack}, {Rozanov}, \& {Unruh}}]{shapiro14}
{Shapiro}, A.~I., {Solanki}, S.~K., {Krivova}, N.~A., {et~al.} 2014, \aap, 569,
  A38

\bibitem[{{Shapiro} {et~al.}(2015){Shapiro}, {Solanki}, {Krivova}, {Tagirov},
  \& {Schmutz}}]{shapiro15}
{Shapiro}, A.~I., {Solanki}, S.~K., {Krivova}, N.~A., {Tagirov}, R.~V., \&
  {Schmutz}, W.~K. 2015, \aap, 581, A116

\bibitem[{{Shapiro} {et~al.}(2016){Shapiro}, {Solanki}, {Krivova}, {Yeo}, \&
  {Schmutz}}]{shapiro16}
{Shapiro}, A.~I., {Solanki}, S.~K., {Krivova}, N.~A., {Yeo}, K.~L., \&
  {Schmutz}, W.~K. 2016, \aap, 589, A46

\bibitem[{{Vaughan} \& {Preston}(1980)}]{vaughan80}
{Vaughan}, A.~H. \& {Preston}, G.~W. 1980, \pasp, 92, 385

\bibitem[{{Witzke} {et~al.}(2018){Witzke}, {Shapiro}, {Solanki}, {Krivova}, \&
  {Schmutz}}]{witzke18}
{Witzke}, V., {Shapiro}, A.~I., {Solanki}, S.~K., {Krivova}, N.~A., \&
  {Schmutz}, W. 2018, ArXiv e-prints

\end{thebibliography}

%%%%%%%%%%%%%%%%%%%%%%%%%%%%%%%%%%%%%%%%%%%%%%%%%%%%%
%%%%%%%%%%%%%%%%%%%%%%%%%%%%%%%%%%%%%%%%%%%%%%%%%%%%%%
%\begin{appendix}

%\end{appendix}

\end{document}